\documentclass[twocolumn]{aastex63}
\usepackage{hyperref}
\usepackage{color}
\usepackage{amssymb}
\usepackage{amsmath}
\usepackage[ruled,linesnumbered]{algorithm2e}
\usepackage{footmisc}
\usepackage{ulem}

\newcommand\cI{\mathcal{I}}
\received{January 1, 0000}
\revised{January 1, 0000}
\accepted{January 1, 0000}
\submitjournal{ApJ}

\shorttitle{IFU Ia DTD}
\shortauthors{Chen et al.}
\begin{document}

\title{Constraining Type Ia Supernova Delay Time with Spatially Resolved Star Formation Histories}
\correspondingauthor{Lifan Wang}
\email{lifan@tamu.edu}

\author{Xingzhuo Chen}
\affiliation{George P. and Cynthia Woods Mitchell Institute for Fundamental Physics \& Astronomy, \\
Texas A. \& M. University, Department of Physics and Astronomy, 4242 TAMU, College Station, TX 77843, USA}
\affiliation{Purple Mountain Observatory, Nanjing 210008, China}

\author{Lei Hu}
\affiliation{Purple Mountain Observatory, Nanjing 210008, China}

\author{Lifan Wang}
\affiliation{George P. and Cynthia Woods Mitchell Institute for Fundamental Physics \& Astronomy, \\
Texas A. \& M. University, Department of Physics and Astronomy, 4242 TAMU, College Station, TX 77843, USA}

\begin{abstract}
    We present the delay time distribution (DTD) estimates of Type Ia supernovae (SNe~Ia) using  spatially resolved SN~Ia host galaxy spectra from MUSE and MaNGA. 
    By employing a grouping algorithm based on k-means and earth mover's distances (EMD), we separated the host galaxy star formation histories (SFHs) into spatially distinct regions and used maximum likelihood method to constrain the DTD of SNe Ia progenitors. 
    When a power-law model of the form $DTD(t)\propto t^{s} (t>\tau)$ is used, we found an SN rate decay slope $s=-1.41^{+0.32}_{-0.33}$ and a delay time $\tau=120^{+142}_{-83} Myr$ . 
    Moreover, we tested other DTD models such as a broken power law model and a two-component power law model, and found no statistically significant support to these alternative models. 
\end{abstract}


\section{Introduction}

Type Ia supernovae (SNe~Ia) are produced by the explosion of white dwarfs (WDs) in binary systemts  \citep[see e.g.,][for reviews]{Maguire2016Handbook,Branch:Wheeler2017suex.book.....B}. 
However, the configuration of the binary system remains unknown, and there are two leading scenarios. 
In the single-degenerate (SD) scenario, the progenitor WD accretes matter from a nondegenerate companion star to reach a critical mass $\sim 1.37 M_{\odot}$ before SN explosion \citep{WhelanIben1973SD,Nomoto1982SD}. 
In the double-degenerate (DD) scenario, the explosion is triggered by the merging of two WDs \citep{Webbink1984DD,Iben1984DD}. 

The delay time distribution (DTD), which describes the SN~Ia rate per unit stellar mass as a function of the time after a burst of star formation activity, is scrutinized in our research to investigate the SD and DD scenarios and the observed SN event statistics. 
In binary population synthesis (BPS) simulations of carbon-oxygen white dwarf (CO WD) and helium white dwarf (He WD) merger \citep{Meng2015COHe,LiuD2017WDHe}, the delay time (denoted as $\tau$) from the star formation activity to the first SN~Ia is typically  $10^{8.5} \sim 10^{9}$ years, and the SN rate follows a $t^{-1}$ decay with time. 
Similarly, the BPS simulations of CO WD and CO WD also show a $t^{-1}$ relation, but the delay time of the first SN~Ia is around $10^{8}\sim 10^{8.5}$ years \citep{LiuD2017WDHe,Chen2012DD}. 
In contrast, simulations for the SD scenario show different delay times for different channels. 
\citet{Claeys2014SDDD} and \citet{Wang2015SuperEdd} simulated the CO WD and main sequence star (CO WD+MS) channel, and found most of the SNe Ia are produced in $10^8\sim 10^9$ years after star formation. 
\citet{LiuD2019WDRG} simulated the CO WD and red giant star (CO WD+RG) channel, and found the delay time to be $10^{8.6}\sim 10^{8.7}$ years but with a steeper delay time relation than that of the DD scenario. 
Apart from MS or RG serving as the companion star, \citet{Wang2017WDHe} simulated CO WD + He star as SN~Ia progenitor system, which shows a peak of the event rate around $10^8$ years. 
In addition, \citet{Denissenkov2013CONe} proposed a new channel with carbon-oxygen-neon (CONe) WD serving as SNe Ia progenitor. 
In both the CONe WD+He star channel \citep{Wang2014WDHe} and CONe WD+MS channel \citep{Meng2014CONeWS}, the delay times are $10^{7.5}\sim 10^{9}$ years and do not show the $t^{-1}$ decaying rate. 
Although different BPS simulations of DD scenario show SN rates in agreement with observations (e.g. \citet{toonenMerger,Ruiter2009Rates,LiuD2017WDHe,Claeys2014SDDD}), the observed SN population can still originate from a combination of multiple channels \citep{Nelemans2013ManyTheory}. 

A few major observational measurements of the DTDs are provided by the SN~Ia rates in different redshift bins in galaxy clusters \citep[e.g.,][]{Matan2018z1d75} and from large untargeted supernova survey projects \citep[e.g.,][]{Madgwick2003SDSS,Graus2013SloanDTD,Frohmaier2019PTF,Perrett2012SNLS,Rodney2014Candels,Heringer2019DTD}. 
Alternatively, SN remnants can be used to measure the SN rate \citep{Maoz2010SNRMagCloud}. 
Although the measured results show a $t^{-1}$ SN rate decay which is consistent with the DD scenario, the exact parameters of the relation are strongly dependent on  the details of the cosmic star formation history (CSFH) \citep{galyam2004ObsDtd}. 
Furthermore, due to the incompleteness of high redshift SNe Ia discoveries in high redshift SN surveys, the delay time $\tau$ is still not very strongly constrained. 

In quest of an independent estimate of the SN~Ia delay time, \citet{Maoz2012Sloan2} used the host galaxy stellar population as a proxy to estimate the DTD assuming that SN progenitors share the same formation history with the other stars in the host galaxy. 
Based on such an assumption, \citet{Tyler2020Iax} utilized Hubble Space Telescope (HST) to directly observe the stars close to 9 type Iax SNe and constrained the delay time using their nearby stellar ages as proxies. 
Furthermore, \citet{Panther2019SN1991bg} utilized Integral Field Unit (IFU) facilities to acquire the galaxy spectra of 17 SN~1991bg-like SNe at the sites of the SN explosions, calculated the stellar populations within $\sim 1\ kpc$ of the SNe, and concluded that SN~1991bg-like SNe originate from an older stellar population than normal SNe Ia. 
Notably, as core-collapse SNe originate from younger star populations, their local star formation histories (SFHs) show distinctively younger stellar populations than those for SNe~Ia, and set strong constraints on the ages of Type II, Ib, Ic, IIb and IIn SN progenitors \citep{Kuncarayakti2018CC}. 
In addition to the correlation between the SN rates and the local SFHs, \cite{Galbany2017HII} proposed to use the distance between the SNe and \ion{H}{2} regions as a SN progenitor age indicator. 
The SN host galaxy star forming region \citep{Galbany2014CalifaI}, SN environmental metallicity \citep{Galbany2016CalifaII}, galaxy velocity field \citep{Zhou2019Manga}, are also discussed for their potential influences on the SN~Ia rate. 

Our research uses the spectra at the SN coordinates as the SN progenitor age indicator. 
We introduced two additional assumptions in our analysis: (1) There is no bias against any Types of SN host galaxies in any SN survey projects; (2) The group of stars at the site of the SN explosion statistically exhibits a higher probability of producing an SN at present time than the other groups of the stars in the galaxies. 
To quantify the probability differences among different groups of stars in the SN host galaxies, we develop a novel algorithm to spatially separate the SFHs of a host galaxy into different subgroups based on their SFH profiles. 
We use maximum likelihood method to constrain the DTD model parameters, which takes into account of the relations between different groups of stars and the SN events. 

In Section \ref{sec:data}, we introduce our SN host galaxy sample selection criteria and the calculation of the SFHs. 
In Section \ref{sec:algorithm}, we present the algorithm on the separation of SFHs of the host galaxies spatially into subgroups and the maximum likelihood estimations of the DTD models. 
In Section \ref{sec:results}, we show the results on the constraints on the DTD model parameters. 
Conclusions and discussions are given in Section \ref{sec:conclusion}. 

\section{Data Reduction}\label{sec:data}

\subsection{Sample Selection}

Most of the data used in our research are taken by the Multi-Unit Spectrograph Explorer (MUSE), an IFU facility mounted on the 8.4 meter telescope Yepun (UT4) at Very Large Telescopes (VLT). 
We use all publicly available SNe~Ia listed on Transient Name Server (TNS) \footnote{\label{note1}\href{https://wis-tns.weizmann.ac.il}{https://wis-tns.weizmann.ac.il}}, and cross-match with available MUSE data. 
For the matched SN-host pairs, the following selection criteria are applied: 

\begin{itemize}
    \item 
    The time difference between SN discovery and IFU observation is at least 100 days to avoid SN light from contaminating the host galaxy's spectra. 
    \item The spectral Type of the SN is normal SN~Ia according to both TNS\footref{note1} and SIMBAD\footnote{\href{http://simbad.u-strasbg.fr/simbad/}{http://simbad.u-strasbg.fr/simbad/}}. 
    \item The spatial area of the host galaxy is covered by more than 70\% by the MUSE data. 
    \item After binning the spectral data cube by 3$\times$3 spatial pixels, the signal-to-noise ratio (SNR) at the position of the SN is at least 0.8 per \AA. 
    \item The host galaxy has at least 200 spatial pixels with $SNR\ >\ 0.8$ 
     per \AA\ in the data spatially binned by 3$\times$3. 
\end{itemize}

\begin{figure}
    \centering
    \plotone{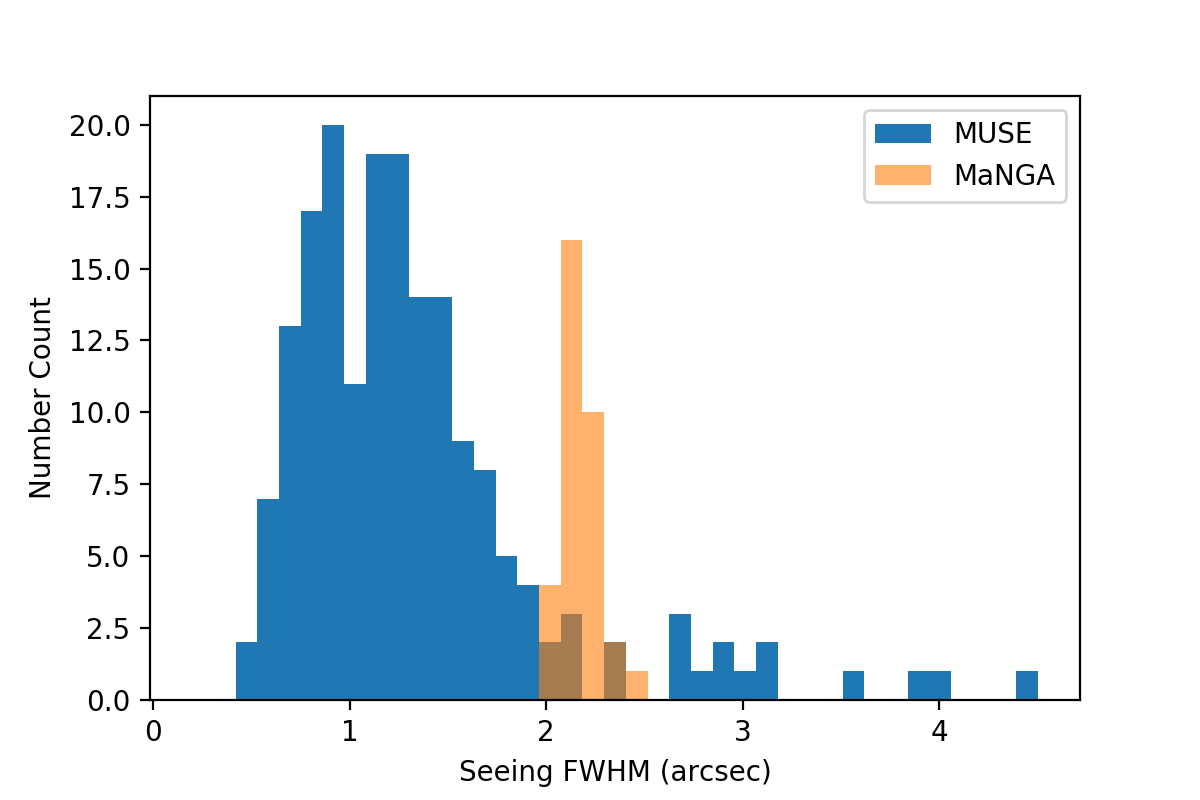}
    \caption{The histogram of the seeings of the data used in our research. 
    Blue columns are from the MUSE data set, orange columns are from the MaNGA data set. }
    \label{fig:seeingHist}
\end{figure}

We found 100 SNe Ia in 96 host galaxies (4 galaxies host 2 SNe in each) satisfying the above criteria, they were observed by 37 VLT observation programs.
The list of all SNe and host galaxies are shown in Appendix \ref{sec:snlist}. 
Among all these observations, four of them (SN~2011iv, SN~2011is, SN~2009ev, SN~1992A) 
have employed the extended wavelength coverage from 465 nm to 930 nm using the {\tt\string WFM-NOAO-E} instrumentation mode, one (SN~2000do) has employed the 
adaptive optics (AO) technique using the {\tt\string WFM-AO-N} instrumentation mode, and the rest are observed in natural seeing with a nominal wavelength range from 480 nm to 930 nm using the {\tt\string WFM-NOAO-N} instrumentation mode. 
The spectral resolving power for both extended wavelength coverage and nominal coverage varies with wavelength, with resolving power $R=2000-4000$, and all the observations used the wide field mode ({\tt\string WFM}) with 1 square arcminute Field of View (FoV). 
In Figure~\ref{fig:seeingHist}, we show a histogram of the seeing distribution. 

We downloaded the data product, which were reduced by the MUSE data reduction pipeline \citep{musedrp}, from ESO Science Archive\footnote{\href{http://archive.eso.org/}{http://archive.eso.org/}}. 
During the multi-year observation period, the version of the data reduction pipeline had changed from {\tt\string v1.4} to {\tt\string v2.8} and we did not attempt to reduce the data again with the latest version of the pipeline. 
Moreover, some data cubes are stacks of multiple observations, which makes the spatial coverage of some data cubes larger than the FoV of the instruments. 

In addition to the MUSE data, the Mapping Nearby Galaxies at APO \citep[MaNGA,][]{manga}, is a spectroscopic survey of galaxies using the 2.5m telescope at Apache Point Observatory (APO) which also provides observations of SN host galaxies. 
The survey has a hexagonal FoV with a diameter varying from 12 arcseconds to 32 arcseconds covering the wavelength range $3600-10300$ \AA\ with a resolving power $R\ \sim\  2000$ and a spatial pixel scale of 0.5 arcseconds. 
We applied the same selection criteria (except binning of the spectral data cube) as used for the MUSE data set, and found 30 SN~Ia host galaxies among 4824 galaxies in SDSS-DR15 \citep{sdssdr15}, the information on the SNe and host galaxies is tabulated in Appendix \ref{sec:snlist}. 
The fully-reduced data cubes, downloaded via {\tt\string marvin} \citep{marvin}, are used to calculate the spatially-resolved SFHs. 

\subsection{Spatially Resolved SFH Calculation}\label{sec:sfh}

As mentioned above, the original MUSE data cubes were binned by 3$\times$3 spatial pixels to increase the SNR of each spectral element. 
The software Source Extractor \citep{sextractor} is used to identify the sky background close to the host galaxy and filter out the foreground stars. 
The pixels with spectral SNR above 0.8 per \AA\ of the SN host galaxies were used to calculate the SFHs. 
Note that some recent studies (e.g. \citep{museppxfearly}) on the SFH with MUSE data employed Voronoi tessellation to build adaptive grids to spatially bin the spectra with lower SNRs. 
We didn't adopt this method because it will mutilate spatial information which is important in our research. 
The MaNGA data cubes were not binned, and we use the mask provided in {\tt\string marvin} to remove foreground stars and bad pixels. 
In all of the MaNGA survey targets, the galaxy covers the entire or most part of the FoV, and the foreground stars are identified in {\tt\string marvin} program, so we didn't use Source Extractor on these galaxies. 

Penalized Pixel-Fitting ({\tt\string ppxf}) \citep{ppxf2017}, which solves for a linear combination of simple stellar population (SSP) models from a stellar spectral library to fit the observed spectra, is used in our research to calculate the host galaxies' SFHs. 
We use the {\tt\string E-MILES} stellar spectral library, and assume Padova stellar evolutionary tracks and isochrones \citep{padova}, \citet{kroupa} initial mass function (IMF), and [$\alpha$/Fe]\ =\ 0 for all the galaxies. 
The model spectral library consists of 50 age grids ranging from 
$63.10$ Myr to $17.78$ Gyr and 7 metallicity grids ranging from $[M/H]=-2.32 $ to $[M/H]=0.22 $. 
We use a 10 order multiplicative polynomial and no additive polynomial to fit the spectral continuum in the spectral fitting process. 

In {\tt\string ppxf}, the smoothness of the SFH is controlled by the regularization parameter {\tt\string regul}. 
We adopt the method introduced in \citet{atlasppxf} and \citet{virgoppxf} to modify the {\tt\string regul} parameter. 
For the SFH calculation of one spectrum, we firstly calculate a series of fitting spectra with {\tt\string regul} changing from $2^{10}$ to $2^{-8}$ as a geometric sequence with common ratio $1/2$. 
Fitting spectra with $regul=0$ is also calculated as the unregularized case. 
Then, we multiply the flux level and set the $\chi^2(regul=0)/N$ statistic for {\tt\string regul}=0 model to be 1, where $N$ is the number of pixels for the observed spectra. 
Finally, the {\tt\string regul} value when $\chi^2(regul)-\chi^2(regul=0)$ of the corresponding fitting is equal or close to $\sqrt{2N}$ is chosen for the spectral fitting, as suggested in \citet{press}. 
Considering there are 96 data cubes and each data cube typically consists of $\sim~3000$ spectra of the SN host galaxy, searching for the optimal {\tt\string regul} parameter for each individual spectrum is computationally prohibitive. 
Accordingly, for each galaxy we only search for {\tt\string regul} using the spectrum with the highest SNR and then apply the {\tt\string regul} value to the SFH calculation of the entire data cube. 
Note that while the intrinsic SFHs may be stochastic and discrete, the smoothed SFHs from our method may not precisely be the true SFHs of the galaxy. 
Smoothing can reduce the degeneracy between age and metallicity from bin to bin and enable the differential comparison of SFHs within each galaxy. 

\section{Methodology}\label{sec:algorithm}

In this study, a galaxy is spatially separated into stellar groups based on the SFHs we have deduced. 
Accordingly, we apply theoretical DTD models to calculate the SN rates of the stellar groups. 
Subsequently, we use maximum likelihood method to derive the optimal parameters of the DTD models to maximize the SN rates of the SN-related groups and minimize the SN rates of the stellar groups unrelated to the SNe Ia for all of the selected galaxies. 
In Section \ref{sec:emd}, we describe the Earth Mover's Distance (EMD) and show how that can be used as a method to calculate the difference between two SFHs. 
In Section \ref{sec:kmeans}, we present our algorithm of spatially separating a galaxy into different groups. 
In Section \ref{sec:dtd}, we present the four DTD models used in our research. 
In Section \ref{sec:likelihood}, we present the likelihood function we have used for the maximize likelihood estimation. 

\subsection{Earth Mover's Distance}\label{sec:emd}

Earth mover's distance (EMD) is developed to evaluate the similarity among distributions and has been widely used in image recognition \citep{emd} and deep-learning \citep{wgan}. 
By definition, EMD measures the minimum amount of work required to change one distribution into the other. 

For $P=\{(x_1, p_1),..., (x_i, p_i), ..., (x_m, p_m) \}$ and $Q=\{(y_1, q_1), ... ,(y_j, q_j), ... ,(y_n, q_n) \}$ as two distributions, where $(x_i,y_j)$ are the centers of data groups $(i,j)$ and $(p_i,q_j)$ are the probabilities of the groups, we can define a flow $F=[f_{ij}]$ between $x$ and $y$ which represents moving the probability in $x_i$ to $y_j$. 
The flow is a feasible flow between $P$ and $Q$ when: 

\begin{eqnarray}
    f_{ij}\geq 0,\ 1\leq i \leq m, 1\leq j \leq n \\ 
    \sum_{j=1}^m f_{ij} \leq p_i,\ 1\leq i \leq m \\ 
    \sum_{i=1}^n f_{ij} \leq q_j,\ 1\leq j \leq n \\ 
    \sum_{i=1}^m \sum_{j=1}^n f_{ij} = min\left(\sum_{i=1}^m p_i, \sum_{j=1}^n q_j\right). 
\end{eqnarray}

For a feasible flow $F(x,y)$, the work done by the flow in matching $P$ and $Q$ distribution is: 

\begin{equation}
    W(F,P,Q)=\sum_{i=1}^{m}\sum_{j=1}^n f_{ij}d_{ij},
\end{equation}
where $d_{ij}$ is the distance between $x_i$ and $y_j$. 
When a flow is an optimal flow $F^*(P,Q)=[f^*_{ij}](P,Q)$, which minimizes $W(F,P,Q)$, the EMD is defined as the work normalized by the optimal flow: 

\begin{equation}
    EMD(P,Q)=\frac{\sum_{i=1}^{m}\sum_{j=1}^n f^*_{ij}d_{ij}}{\sum_{i=1}^m \sum_{j=1}^n f^*_{ij}}. 
\end{equation}

As mentioned in Section \ref{sec:sfh}, the spectral isochrone \citep{padova} used in our research is evenly sampled in $log_{10}(t)$ space with 50 grids between $63.10$ Myr and $17.78$ Gyr, therefore we use $log_{10}(t_i)-log_{10}(t_j)$ for the distance $d_{ij}$. 
Moreover, all the SFHs are normalized so that $\sum_{i=1}^{50} p_{i}=\sum_{j=1}^{50} q_{j}=1 $ and $EMD(P,Q)=\sum_{i=1}^{50}\sum_{j=1}^{50} f_{ij}d_{ij}$. 

\subsection{K-means Clustering}\label{sec:kmeans}

In the conventional k-means clustering algorithm, the distances are measured in Euclidean space. 
Given a data set and a target number ($k$) of groups, the k-means algorithm first generates $k$ ``initial" centroids, then assign data points into $k$ groups according to the nearest centroids in Euclidean metric. 
The means are then updated using the centroids of the groups, and data points are iteratively re-assigned into new groups. 

In our algorithm to separate a galaxy into different groups, an SFH profile in a spatial pixel serves as a data point similar to the original k-means clustering algorithm, the mass-averaged SFH of all data points in a group serves as the centroid in the original k-means clustering algorithm, and the EMD between an SFH profile in a spatial pixel and the mass-averaged SFH replaces the Euclidean distance in the original k-means algorithm. 
The algorithm is shown in Algorithm \ref{algo:kmeans}. 
The objective of this algorithm is to derive a map $\cI(x,y)$ which records the group assignment of the stellar population at the pixel $(x, y)$, with the inputs being the SFH data cube and the total number of groups $K$. 
Different from the conventional k-means algorithm, we introduce an extra operation in which a group with only one pixel will be eliminated and merged with its closest group. 

\begin{algorithm}\label{algo:kmeans}
    \SetAlgoLined
    \SetKwInOut{Input}{input}
    \SetKwInOut{Output}{output}
    \Input{Grouping number $K$}
    \Input{SFH data cube}
    initialize $\cI(x,y)_0\leftarrow Rand[1,2,...,K]$ \; 
    initialize $i\leftarrow 1$\;
    \While{$\exists\ \cI(x,y)_{i-1} \neq \cI(x,y)_{i}$}{
        \If{$\exists\ n_0\in [1,2,...,K]$, $\cI(x_0,y_0)_i=n_0\ \&\ \forall (x,y)\neq (x_0,y_0),\ \cI(x,y)_i\neq n_0$ }{
            $K\leftarrow K-1$\;
        }
        $MSFH_n \leftarrow$ stellar mass averaged SFH for the n-th group, $n \in [1,2,...,K]$\;
        \For{SFH(x,y) in SFH data cube}{
            $D_n\leftarrow EMD\left(SFH(x,y),MSFH_n\right)$\;
            $\cI(x,y)_{i+1}\leftarrow arg\ min_n\ D_n$\;
            }
        $i\leftarrow i+1$
    }
    \caption{k-Means Based on EMD}
\end{algorithm}

\begin{figure*}[htb!]
    \plottwo{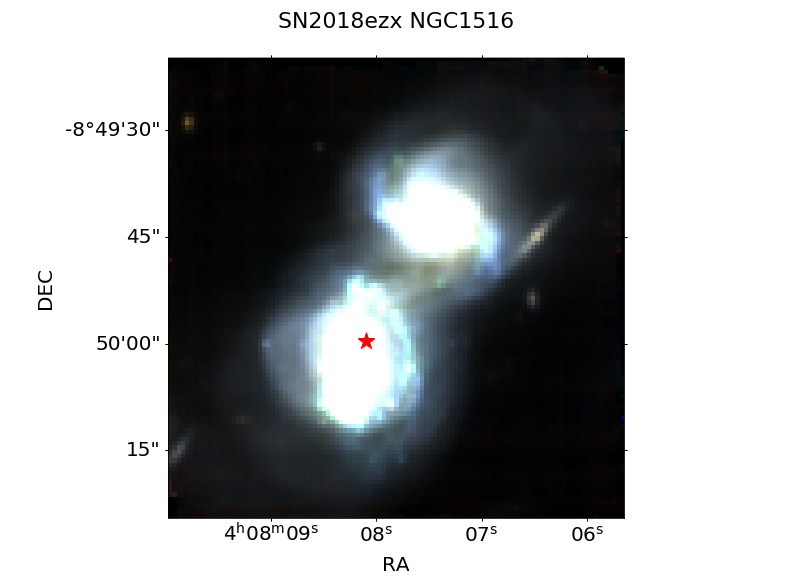}{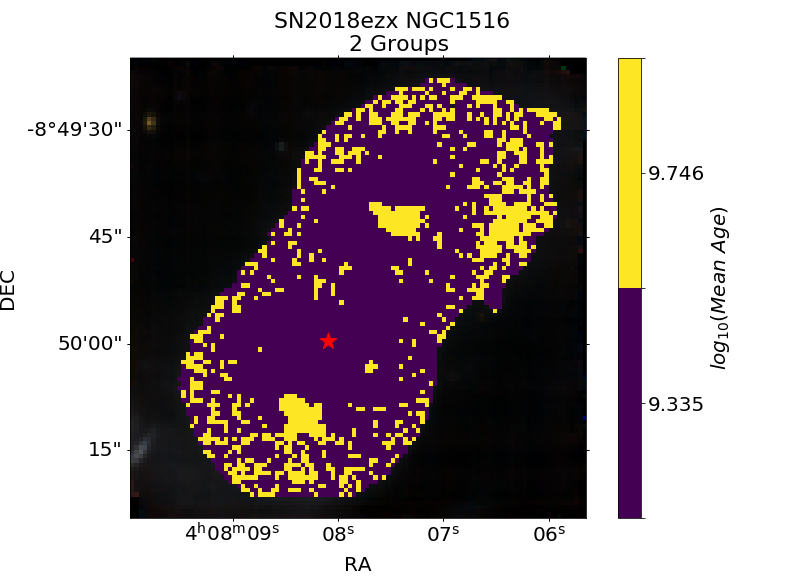}
    \plottwo{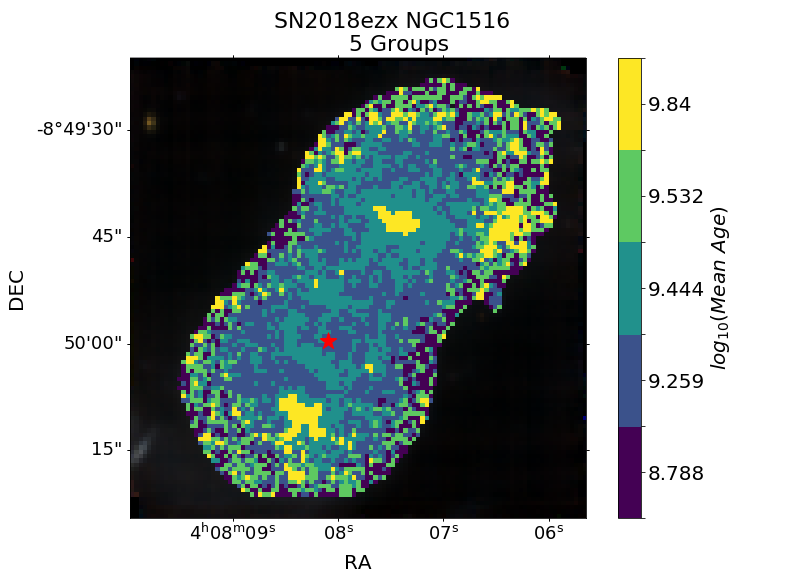}{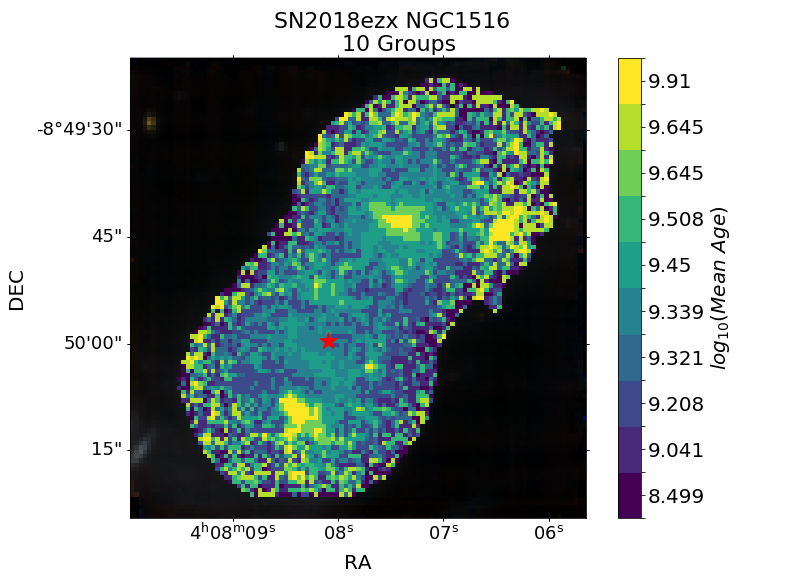}
    \caption{Upper left: Image of NGC~1516, the coordinate of SN2018ezx is marked by the red star. 
    Upper right: The grouping map $\cI(x,y)$ of 2 total groups for NGC~1516. 
    Lower left: The grouping map $\cI(x,y)$ of 5 total groups for NGC~1516. 
    Lower right: The grouping map $\cI(x,y)$ of 10 total groups for NGC~1516. 
    The mean ages of the SFH groups are encoded in different colors and are labeled in the color bars.
}\label{fig:exampleMap}
\end{figure*}

We choose NGC~1516, the host galaxy of SN~2018ezx, to test the algorithm. The SNR of the MUSE observations of NGC~1516 is high, and there are 8682 spatial pixels satisfying our spectral selection limit (\S~\ref{sec:data}). 
It is an interacting galaxy with both galaxies inside the FoV of MUSE, which can potentially have two or more groups of SFHs in the data cube. 
We choose the total number of groups to be 2, 5, and 10 to test the algorithm, and changed the initial $\cI(x,y)_0$ multiple times to test the robustness of the results. 
Typical computational costs are 24 seconds, 98 seconds and 351 seconds for 2, 5 and 10 groups, respectively, using one core of Intel Xeon E5-2670 v2. 
Also, with different random seeds for the initial $\cI(x,y)_0$ maps, the final results are not affected. 
This test verifies that the algorithm can produce stable results for different initial $\cI(x,y)_0$.

In Figure \ref{fig:exampleMap}, we show the final group maps $\cI(x,y)$ of NGC~1516 with group numbers $K=2,5,10$. 
The mean ages of the stellar populations are color encoded in which the ages increase going from purple to yellow. 
The center of the upper-right galaxy and the southern part of the lower-left galaxy are classified as the oldest stellar group in all cases. 
The outskirts of the galaxies show a mixture of young and old SFH groups, and could affect the group of the SN if the SN coordinate is in such a region. 
We surmise this phenomenon could be due to the uncertainties in SFH calculation introduced by the low SNR of the data. 
In Equation \ref{eq:Q} of Section \ref{sec:likelihood}, we will discuss the effect of observational seeing on the SN probability calculation to mitigate this problem. 

\begin{figure}
    \plotone{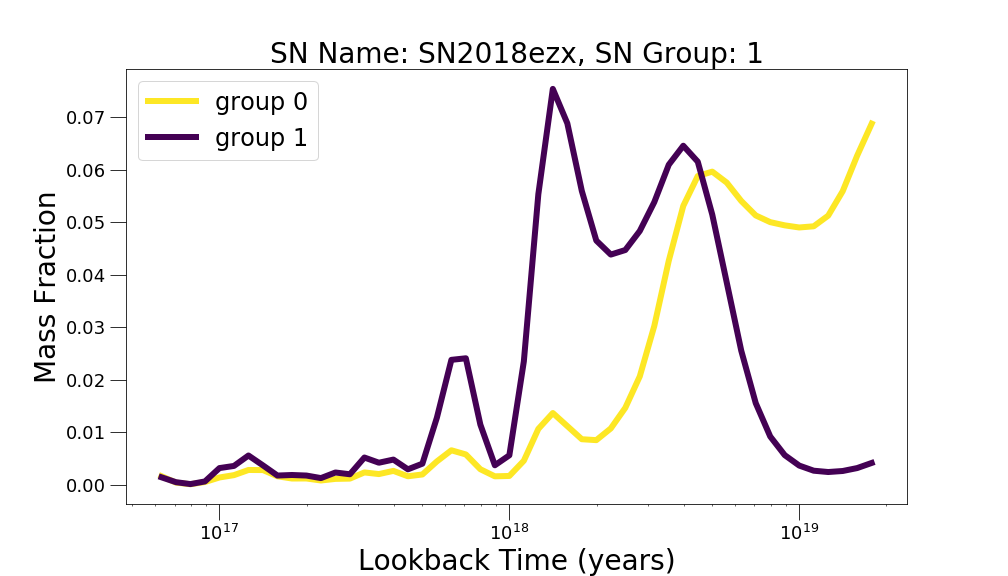}
    \plotone{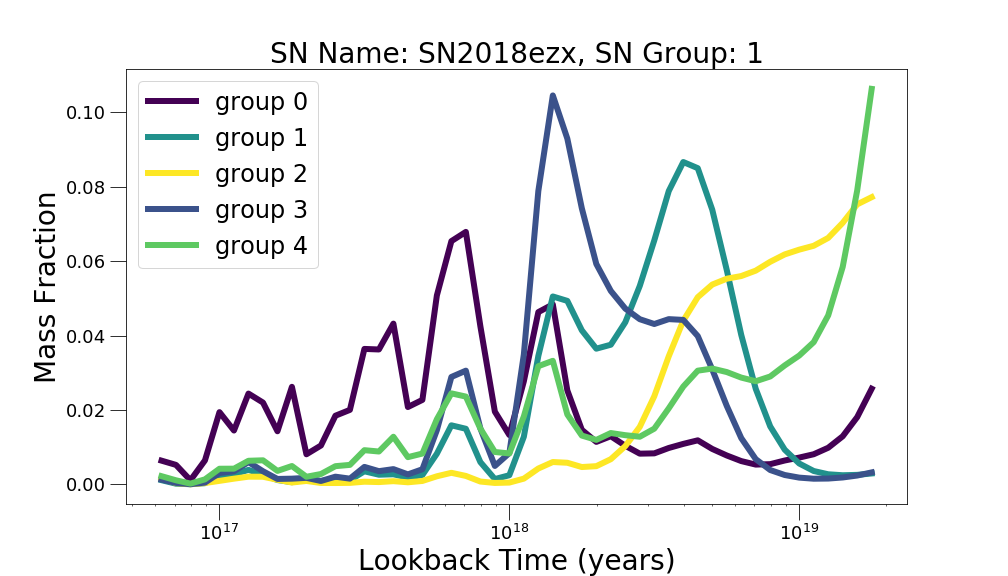}
    \plotone{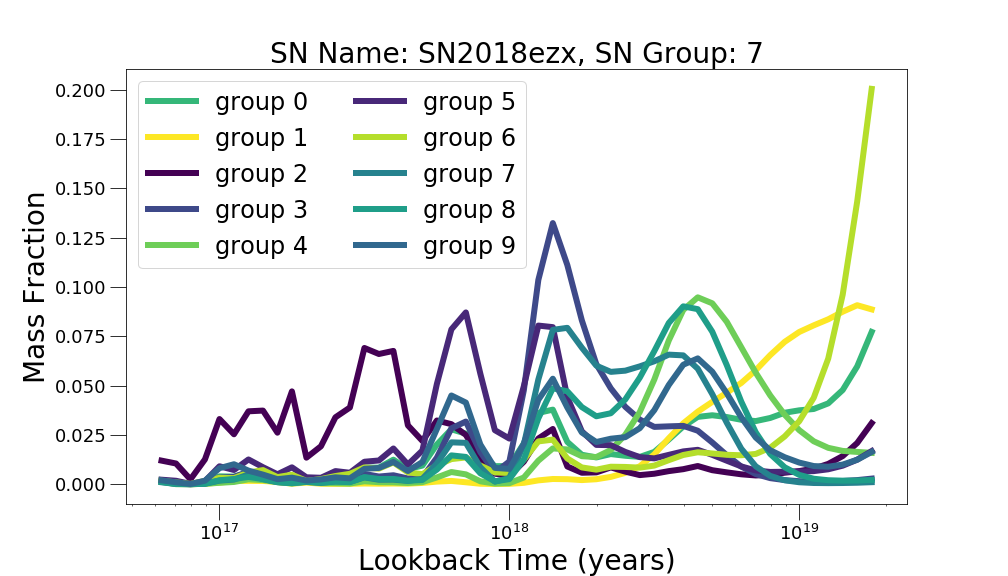}
    \caption{From top to bottom, the mean SFHs of the host galaxy of SN~2018ezx grouped into 2, 5, and 10 total groups, respectively. 
    The groups associated with the SN location are 1, 1, and 7 from top to bottom. 
    The colors of the lines are matched with the colors of the groups in Figure~\ref{fig:exampleMap}. 
    }\label{fig:exampleSFH}
\end{figure}

We present the mass-averaged SFHs of each group in Figure~\ref{fig:exampleSFH}, the ages are encoded in the colors of the curves. 
We notice that for the 2-group and the 5-group separations, the averages of the SN-related groups peak at $10^{9.1}\ yr$; for the separation with 10 groups, most of the pixels close to the SN coordinate belong to group 3 and the SFH of which also shows a peak at $10^{9.1}\ yr$. 

These exercises suggest that the ages of the SN progenitors can be estimated by comparing the SFHs of the host galaxies at the locations of the SNe with those away from them. 
However, not all the SNe host galaxies show such a distinct signal, and maximum likelihood estimation is necessary to estimate the DTD. 

\subsection{Delay Time Distribution Models}\label{sec:dtd}

Delay Time Distribution (DTD) describes the SN rate of a burst of star formation activity after a given time $t$. 
In our research, we use four candidate DTD models. 
The first model (denoted as MDA) is a simple power-law DTD model with two parameters to be constrained. The SN Ia rate evolution with time in MDA is in the following form: 

\begin{equation}
    DTD(t)=\begin{cases}
        0 ,&  t<\tau \\ 
        A\times B\times t^{s}, &  t\geq \tau \\
        
    \end{cases}    
    \label{equation:MDA}
\end{equation}
where $\tau$ is the delay time, $s$ is the slope index, 
$A$ is the normalization factor for the absolute SN rate which can be calibrated by the SN rates derived from SN surveys, $B$ is the normalization factor as defined below in Equation~\ref{equation:NormB}. 
In our calculations, $\tau$ and $s$ are constrained using maximum likelihood estimate. 
In the maximum likelihood estimate, we choose a plain prior for $\log(\tau$) and $s$: $log_{10}(\tau) \in [7.3,10]$; $s\in [-6,0]$. 

The second model (denoted as MDB) is a broken power-law DTD model with four parameters. 
The SN Ia rate is: 

\begin{equation}
    DTD(t)=\begin{cases}
        0 ,&  t<\tau \\ 
        A\times B\times t^{s_1} ,& t_c \geq t\geq \tau \\
        A\times B\times t_c^{s_1-s_2} t^{s_2},& t\geq t_c \\

    \end{cases}
    \label{equation:MDB}
\end{equation}
where the symbols $A$, $B$, and $\tau$ have the same meanings as in Equation~\ref{equation:MDA}, $t_c$ is the critical time, $s_1$ and $s_2$ are the slopes for the two power law components. 
$\tau$, $s_1$, $s_2$, and $t_c$ will be constrained in our calculation. 
In the maximum likelihood estimates, the priors are $log_{10}(\tau) \in [7.3,10]$; $s_1 \in [-6,0]$; $s_2 \in [-6,0]$; $log_{10}(t_c) \in [9.25, 10.1]$. 

The third model (denoted as MDC) is the sum of two power-law relations with the same slope. 
There are four parameters to be constrained. 
The SN Ia rate is given by: 

\begin{equation}
    DTD(t)=\begin{cases}
        0 ,&  t<\tau \\ 
        A\times B\times t^{s} ,& t_c \geq t\geq \tau \\
        A\times B\times r \times t^{s},& t\geq t_c \\

    \end{cases}
\end{equation}
where the symbols $A$, $B$, $s$, and $\tau$ have the same meanings as in Equation~\ref{equation:MDA}, $t_c$ is the critical time, $r$ is the ratio between the two components. 
In the maximum likelihood method, the priors are $log_{10}(\tau) \in [7.3,10]$; $s \in [-6,0]$; $log_{10}(t_c) \in [9.25, 10.1]$; $log_{10}(r)\in [-3,3]$. 

The fourth model (denoted as MDD) is also a combination of two power-law relations, but the slopes are different so the model contains five parameters to be constrained. 
The SN Ia rate is: 

\begin{equation}
    DTD(t)=\begin{cases}
        0 ,&  t<\tau \\ 
        A\times B\times t^{s_1} ,& t_c \geq t\geq \tau \\
        A\times B\times r \times t_c^{s_1-s_2} t^{s_2},& t\geq t_c \\

    \end{cases}
\end{equation}
where $s_1$ and $s_2$ are the slopes for the two components.
In the maximum likelihood estimate, the priors are $log_{10}(\tau) \in [7.3,9]$; $s_1 \in [-6,0]$; $s_2 \in [-6,0]$; $log_{10}(t_c) \in [9.25, 10.1]$; $log_{10}(r)\in [-3,3]$. 

The normalization factors $B$ in all the four DTD models satisfy: 

\begin{equation}
    \int_0^{T_{cosmic}}DTD(t)dt=A,
        \label{equation:NormB}
\end{equation}
where $T_{cosmic}$ is the cosmic age. 
We set $T_{cosmic}=17.78$ Gyr to conform with the SFH population grid used in {\tt\string ppxf}. 
$B$ is effectively a normalization factor such that the coefficient $A$ is directly determined from observed cosmic supernova rates (see Section \ref{sec:csnr}). 


\subsection{Maximum Likelihood Estimate}\label{sec:likelihood}

The probability of finding an SN at position $(x_0, y_0)$ at the present time can be obtained by integrating the contributions from all stars in the host galaxy and through the cosmic times of their evolution,   

\begin{equation}
\label{eq:PSNa}
    \begin{split}
    P_{SN}(x_0,y_0)\propto
    \int_0^{t_{max}} dt'\int\int dxdy\ \Omega(x-x_0,y-y_0) * & \\ R(x,y)  SFH_\cI(t',x,y)*DTD(t'),
    \end{split}
\end{equation}
where $t'$ is the lookback time, $t_{max}$ is the age of the Universe, $R(x,y)$ is a scaling factor that accounts for the total number of stars formed at position $(x,y)$ and is given by the ratio of the observed and the model ppxf spectrum at the position $(x, y)$ , $\Omega(x-x_0, y-y_0)$ is a window function accounting for the probability that the SN at $(x_0, y_0)$ can be related to the stellar groups located at any given $(x, y)$, and $SFH_\cI(t',x,y)$ is the SFH at position $(x,y)$ which is one of the $K$ subgroups as defined in Section~\ref{sec:algorithm}. 

The conditional probability that an SN occurs at position $(x_0, y_0)$ knowing there is an SN from the host galaxy is
\begin{equation}
\label{eq:PSNb}
   \begin{split}
    P_{R}(x_0,y_0) =   \\  
    \frac{P_{SN}(x_0,y_0)}{
    \int_0^{t_{max}} dt'\int\int dxdy R(x,y)  SFH_\cI(t',x,y)*DTD(t').}  \\
   \end{split}
\end{equation}

We adopt a simple Gaussian window function of the form 
\begin{equation}
   \label{eq:Q}
   \begin{split}
    \Omega(x-x_0,y-y_0) =  \\  
    \frac{1}{2\pi \sigma_{D}^2}exp\left( -\frac{1}{2}\left( \frac{x-x_0}{\sigma_{D}}\right)^2 -\frac{1}{2} \left( \frac{y-y_0}{\sigma_{D}}\right)^2  \right),
    \end{split}
\end{equation}
where $\sigma_{D}$ is the Gaussian width whose minimum value is given by the seeing of the observations. 

The joint likelihood function for all of the selected SNe ($N_{SN}$) is: 
\begin{equation}\label{eq:oneLike}
    L_{k}=\prod_{l=1}^{N_{SN}}P_{R,l}\ \ ,
\end{equation}
where $P_{R,l}$ is conditional probability of the $l$-th SN in the sample calculated using Equations~\ref{eq:PSNa}. 
We use the Markov Chain Monte Carlo based code {\tt\string emcee} \citep{emcee} for maximum likelihood estimation of the DTD parameters. 

\section{Results}\label{sec:results}

The MUSE data set is larger than the MaNGA data set and is with much higher quality, we focus our research on the MUSE data, but also provide the results from MaNGA for comparisons. 
In Section \ref{sec:GroupNum}, we show the results of the DTD estimates after applying different numbers of groups $K$ to the MUSE data. 
In Section \ref{sec:mangaresult}, we discuss the DTD estimates using the MaNGA data. 
In Section \ref{sec:comparison}, we apply four different DTD models to the MUSE data and compare the model performances using the Bayesian information criterion (BIC). 

\subsection{The Total Number of SFH Groups}\label{sec:GroupNum}

\begin{figure*}[htb!]
    \minipage{0.33\textwidth}
        \includegraphics[width=\textwidth]{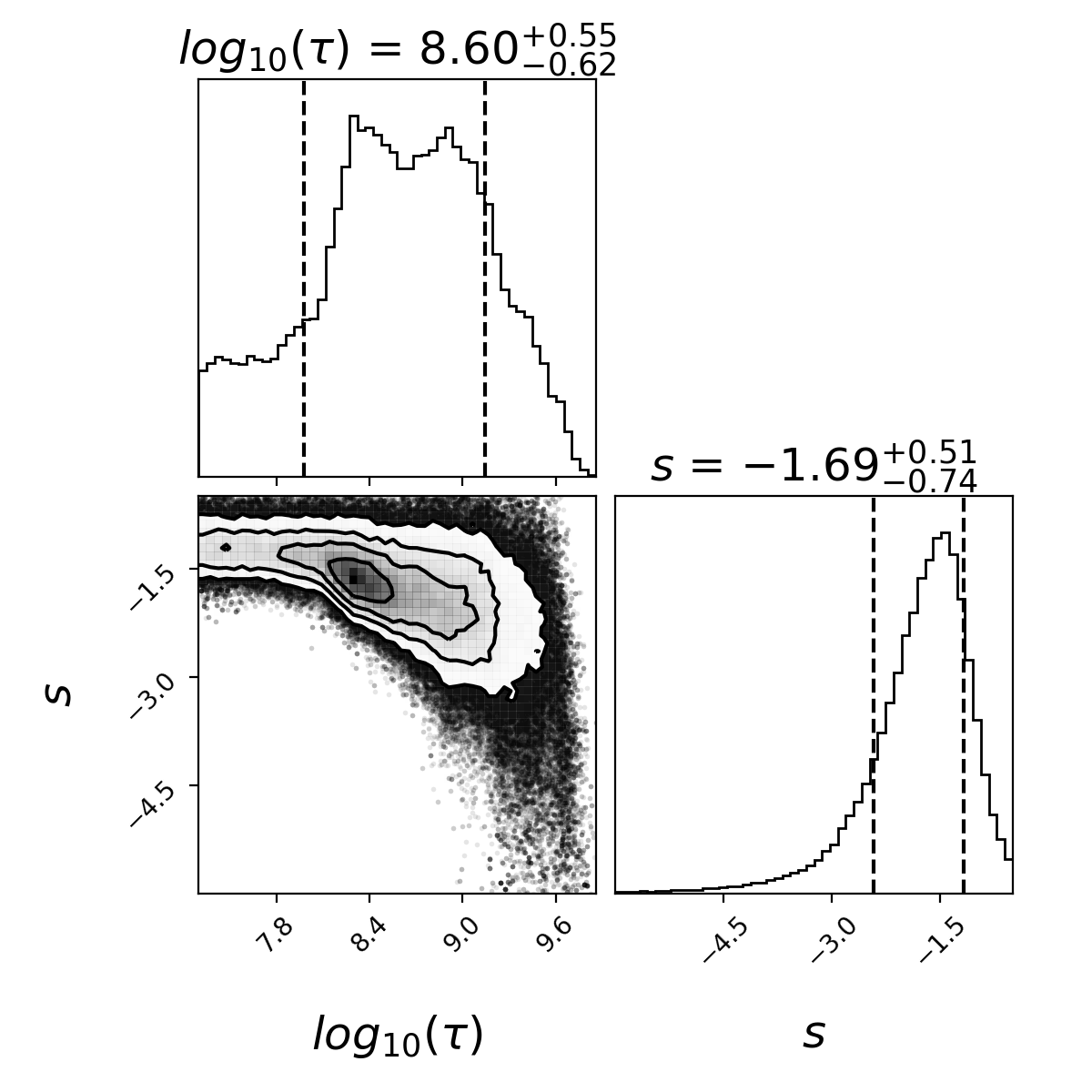}
    \endminipage\hfill
    \minipage{0.33\textwidth}
        \includegraphics[width=\textwidth]{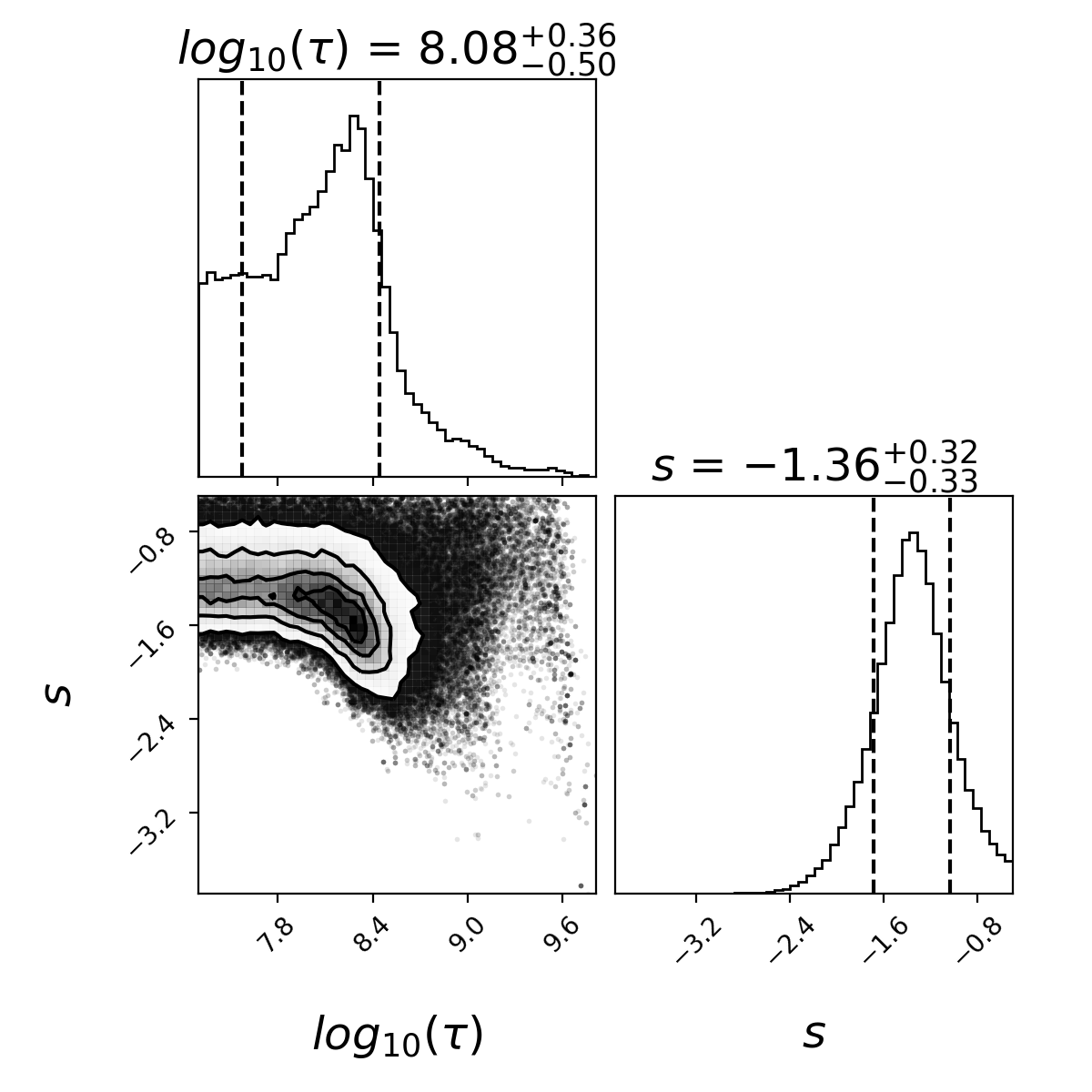}
    \endminipage\hfill
    \minipage{0.33\textwidth}
        \includegraphics[width=\textwidth]{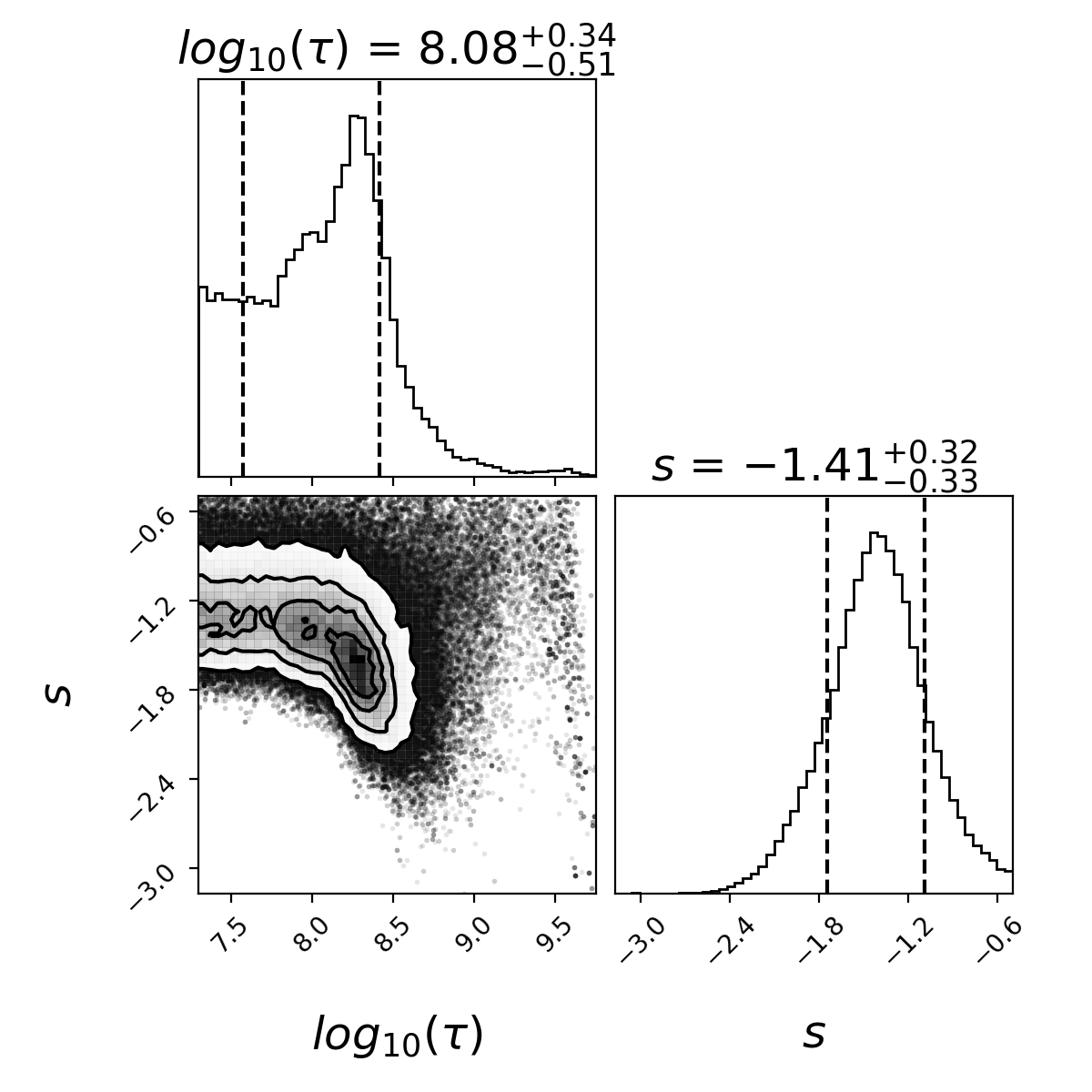}
    \endminipage\hfill
    \caption{From the left to the right, the posterior probability distributions of the MDA model parameters using $k=$2, 7, and 15 groups for the MUSE data. 
    }\label{fig:cornerOne}
\end{figure*}

In this section, we investigate the effect of the number of SFH groups on the constraints of the DTD parameters. 
In Equation \ref{eq:Q}, $\sigma_{D}$ is set to be the seeing profiles of the observations. 
In Figure~\ref{fig:cornerOne}, we present the posterior probability distribution of the MDA model with $K=2,7,15$. 
We notice that the maximum likelihood for all the posterior probability distributions are around $log_{10}(\tau)=8.2$ and $ s=-1.2 $, and the $1\sigma$ intervals for both $\tau$ and $s$ parameters are more tightly constrained with larger values of $K$. 

\begin{figure}[htb!]
    \plotone{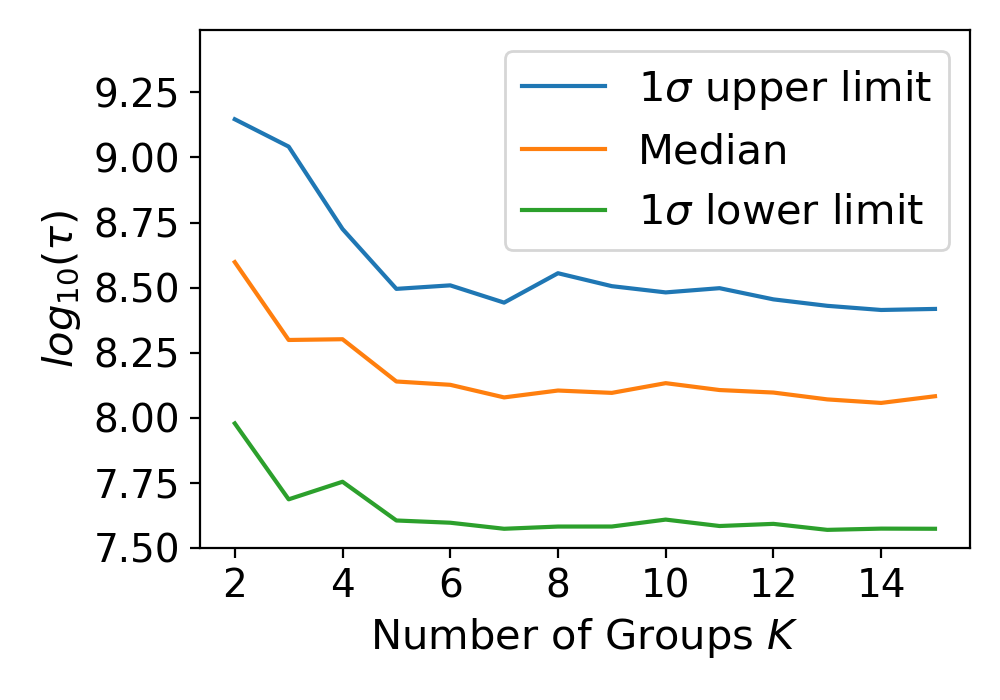}
    \plotone{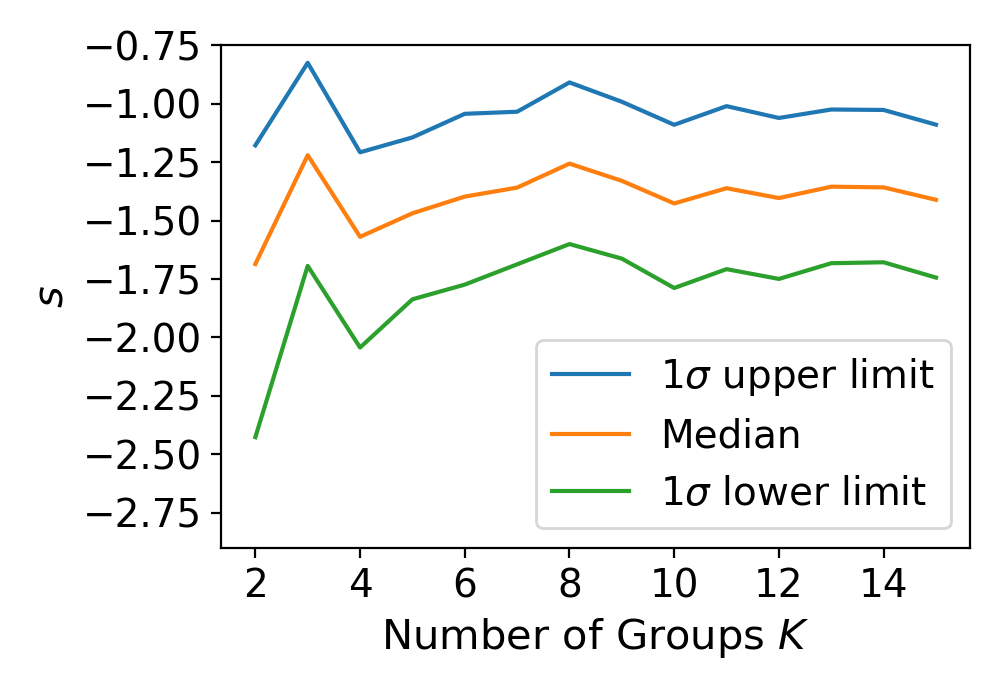}
    \caption{Upper panel: The $1\sigma$ upper limit, median, and $1\sigma$ lower limit of $log_{10}(\tau)$ with different number of groups $K$. 
    Lower panel: The $1\sigma$ upper limit, median, and $1\sigma$ lower limit of $s$ with different number of groups $K$. }\label{fig:Kconstrain}
\end{figure}

In Figure \ref{fig:Kconstrain}, we show the $1\sigma$ limits and the median values of $log_{10}(\tau)$ and $s$ estimated with different grouping numbers. 
The results using $K$ from 2 to 15 are consistent with each other, but with larger fluctuations for $K$ smaller than 4.
It is encouraging that the fluctuations decrease with increasing $K$ values, and converge at the higher end. 
For $K=15$, the MDA model parameter median and $1\sigma$ limit estimates are: $\tau=120^{+142}_{-83}Myr$, $s=-1.41^{+0.32}_{-0.33}$. 

\begin{figure}[htb!]
    \plotone{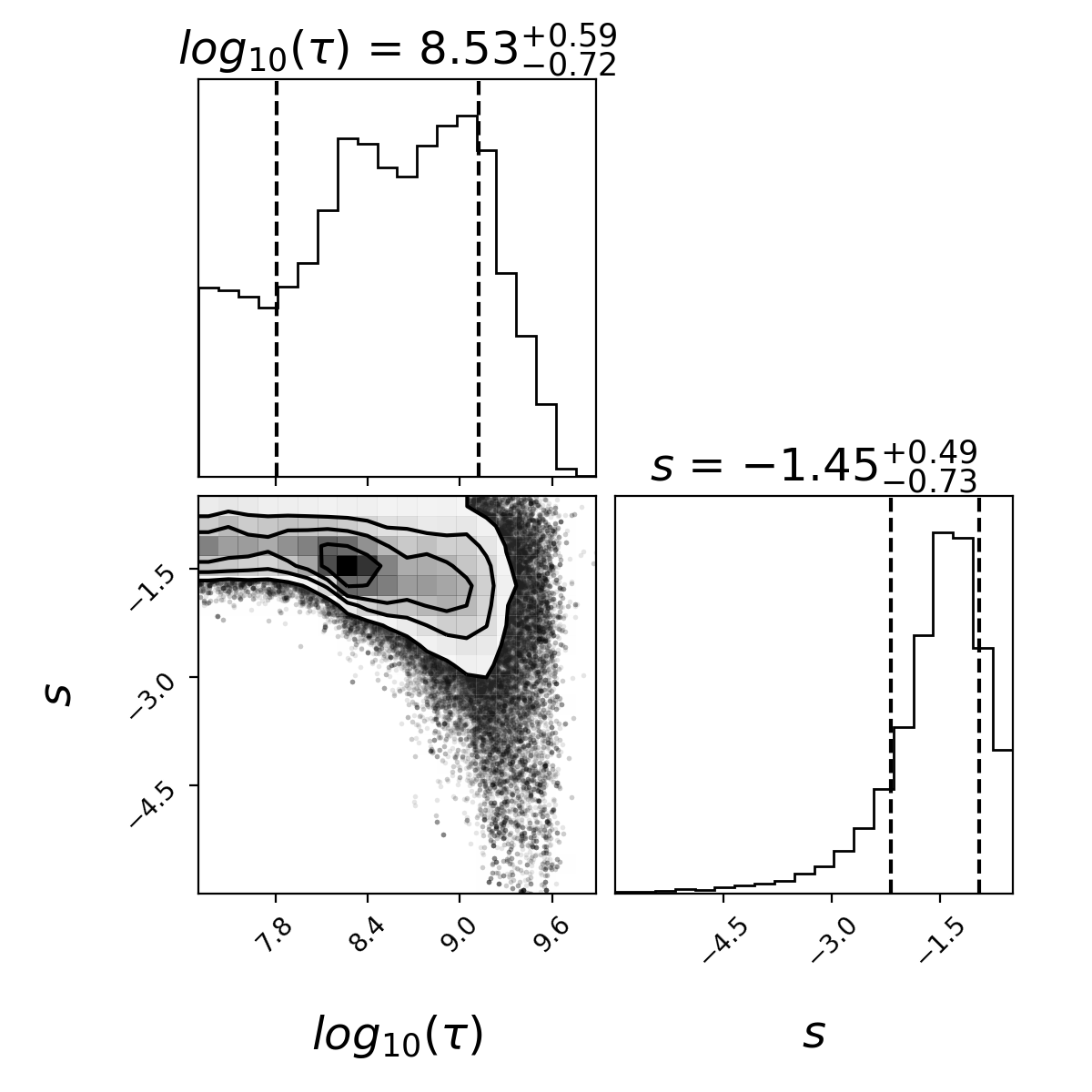}
    \caption{The posterior probability distribution of the MDA model parameters using $K=15$ and using Equation \ref{eq:sigmaD} for $\sigma_{D}$. }\label{fig:5Kpc}
\end{figure}

Assuming the SN progenitor is $\sim 5\ kpc$ from its birthplace, we modify $\sigma_{D}$ in Equation \ref{eq:Q} to be: 

\begin{equation}\label{eq:sigmaD}
    \sigma_{D}=\sqrt{\sigma_{seeing}^2+\sigma_{5kpc}^2}, 
\end{equation}
where $\sigma_{seeing}$ is the Gaussian width of the seeing of the observation, $\sigma_{5kpc}$ is the projected angular distance of 5 kpc in the host galaxy. 
In Figure \ref{fig:5Kpc}, we show the $1\sigma$ limits and the median values of $log_{10}(\tau)$ and $s$  estimated with $K=15$. 
The results are in agreement but show larger uncertainties than those using the observational seeings as $\sigma_D$.

\subsection{Results from MaNGA}\label{sec:mangaresult}

\begin{figure}[htb!]
    \plotone{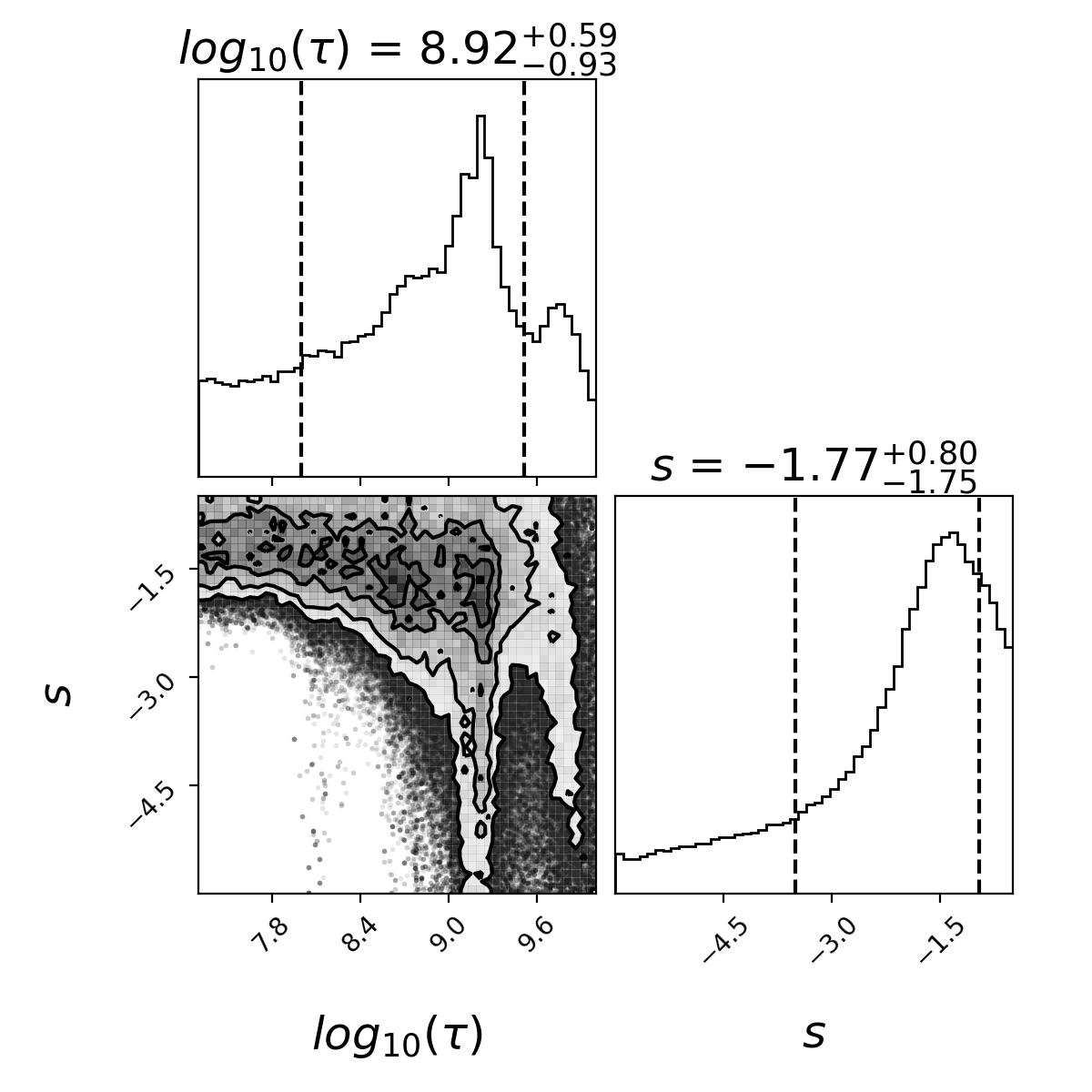}
    \caption{The posterior probability distribution of the MDA model parameters using $K=15$ groups for the MaNGA data.}\label{fig:manga}
\end{figure}

The posterior probability distribution for the DTD model MDA is shown in Figure~\ref{fig:manga}. 
The number of group is set to 15, and $\sigma_{D}$ is the observational seeing. 
We notice the delay time is $832_{-734}^{+240}Myr$ and the slope is $s=-1.77^{+0.80}_{-1.75}$, which is in broad agreement with the values deduced from the MUSE data. 
The MaNGA data give larger uncertainties due to three reasons. 
(1) MaNGA's smaller FoV limits the spectral and SN sample to the center of the galaxy, thus produced a biased DTD estimate. 
(2) MaNGA's seeing is worse than MUSE's, which may introduce more uncertainties in SFH calculation. 
(3) The data set size from MaNGA is smaller than that from MUSE. 

\subsection{Model Comparison}\label{sec:comparison}

\begin{figure}[htb!]
    \plotone{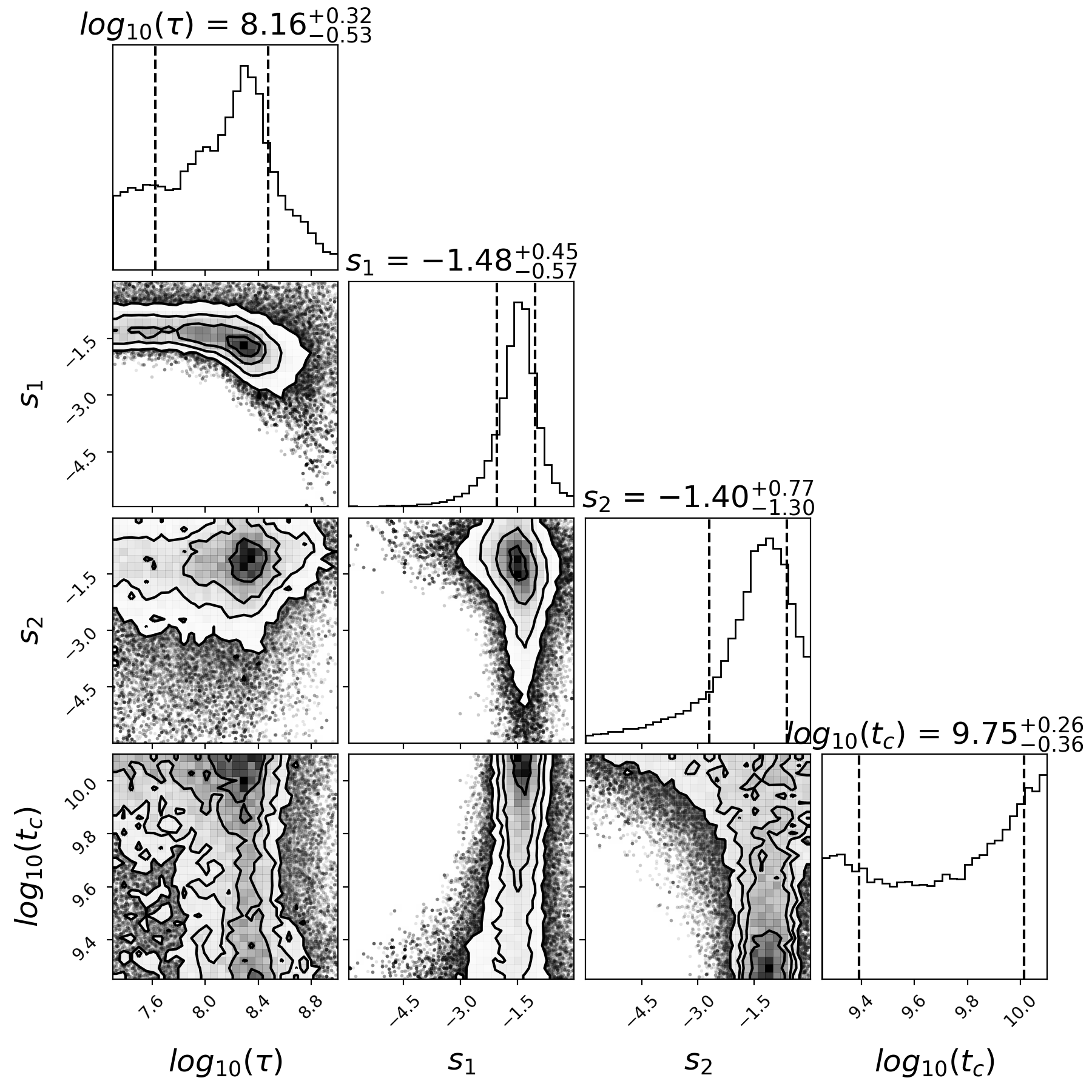}
    \caption{The posterior probability distribution of MDB model parameters using $K=15$ for the MUSE data. }\label{fig:cornerMDB}
\end{figure}

\begin{figure}[htb!]
    \plotone{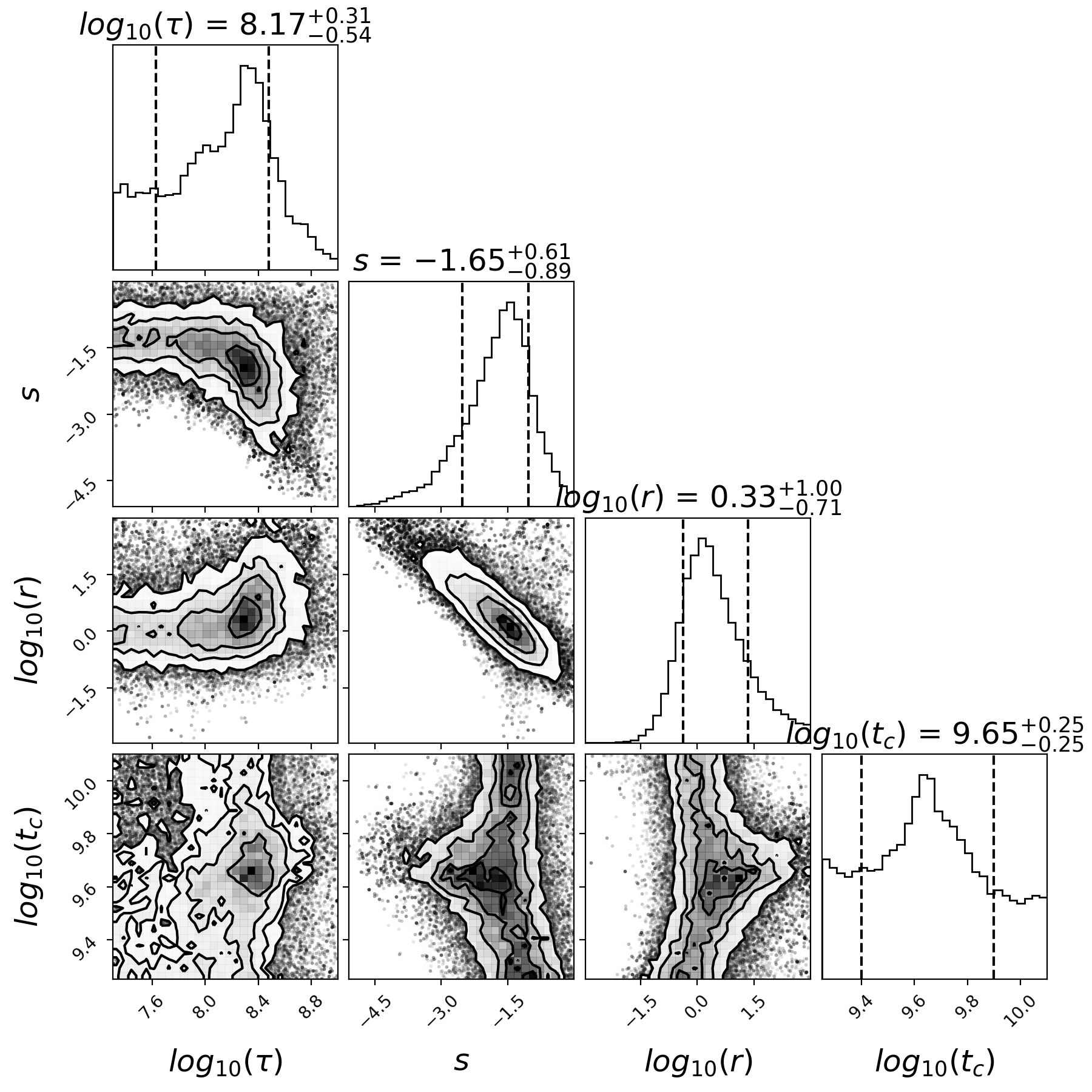}
    \caption{The posterior probability distribution of MDC model parameters using $K=15$ for the MUSE data. }\label{fig:cornerMDC}
\end{figure}

\begin{figure}[htb!]
    \plotone{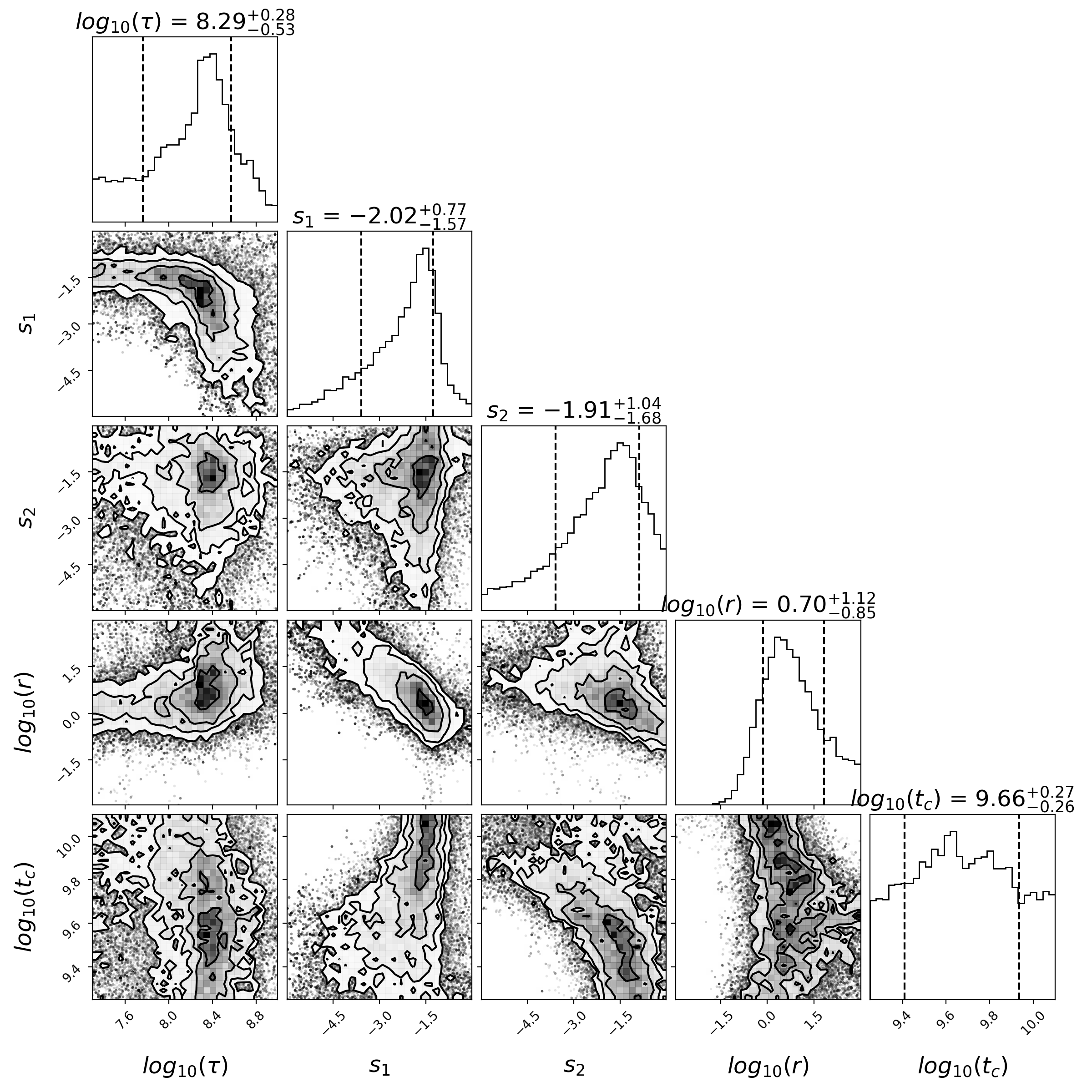}
    \caption{The posterior probability distribution of MDD model parameters using $K=15$ for the MUSE data. }\label{fig:cornerMDD}
\end{figure}

Based on Section \ref{sec:GroupNum}, we use Equation \ref{eq:oneLike} as the likelihood function and set $K=15$ for the parameter constraints of DTD models MDB, MDC and MDD. The results are shown in Figure~\ref{fig:cornerMDB}, Figure~\ref{fig:cornerMDC}, and Figure~\ref{fig:cornerMDD}, respectively. 

Comparing to the results using the MDA model, the parameters in the MDB, MDC and MDD models are less constrained, but still show a maximum likelihood at around $log_{10}(\tau)=8.2$. 
In the posterior probability distribution of MDB, the two slope parameters $s_1$ and $s_2$ are close to the slope $s$ in MDA.
Moreover, the critical time $t_c$ for MDB is close to the age of Universe, which questions the necessity of the second component employed in MDB. 
The parameters in MDC and MDD ($s$ in MDC and $s_1$ and $s_2$ in MDD; $r$ in MDC and MDD; $t_c$ in MDC and MDD) are all in agreement to within the statistical errors. In all these models, $t_c$ was found to be around $10^{9.5}$ years. 
Both MDC and MDD give a value for the ratio $r$ larger than 1, which suggests a population of SNe Ia descend from old stellar populations at around $10^{9.6}$ years albeit with large errors. 

\begin{figure}[htb!]
    \plotone{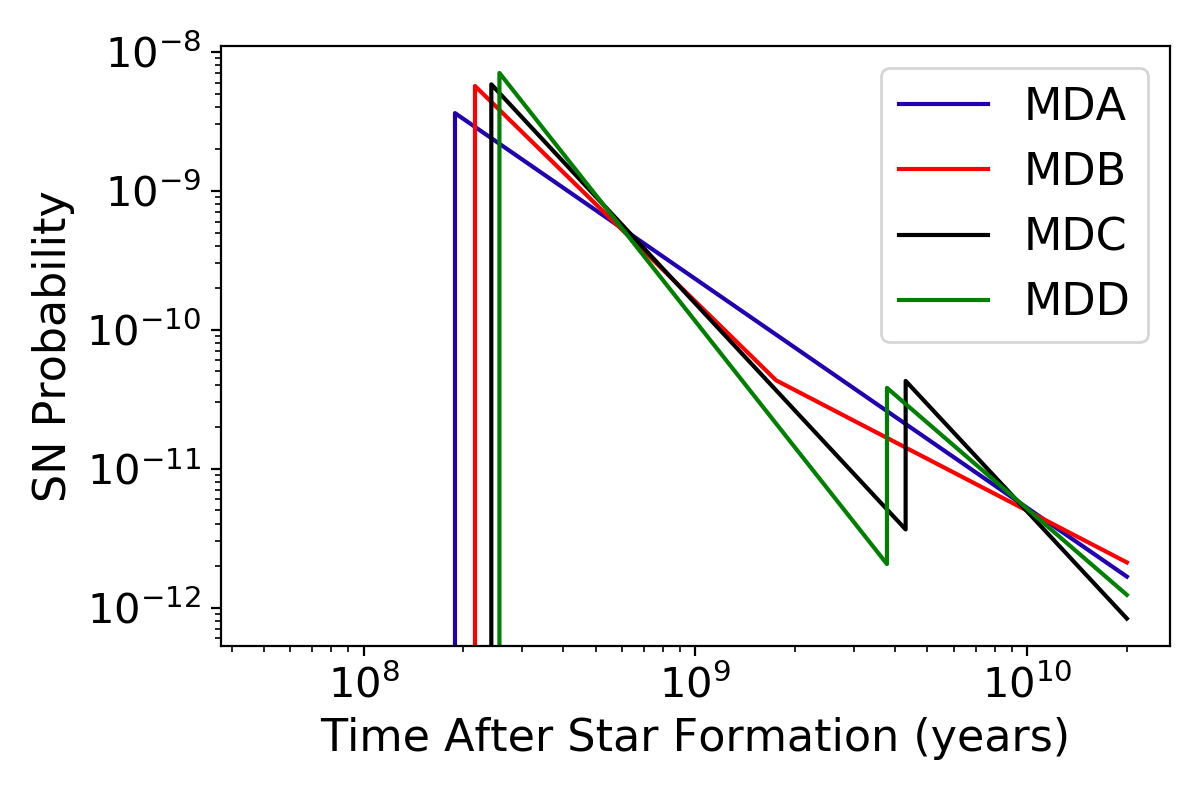}
    \caption{
    The SN Ia rate for the DTD models MDA (ultramarine), MDB (red), MDC (black), and MDD (green). All the DTDs are normalized so that their time integral is equal to 1. The parameters are shown in Table \ref{tab:bic}}\label{fig:DTDcurve}
\end{figure}

To investigate further whether the data can set constraints on progenitor systems with drastically different DTDs, we adopted the Bayesian information criterion (BIC) \citep{Wit2012BIC} to assess the goodness of the fits using the different DTD models. 
The BIC is defined as
\begin{equation}
    BIC=k_{dof}\ ln(n)-2 ln(L),
\end{equation}
where $k_{dof}$ is the degree of freedom, $n$ is the size of data sample, $L$ is the maximized likelihood function. 
In our research, $n=100$ as we have 100 SNe Ia coordinates, $L$ is calculated from Equation~\ref{eq:oneLike}, and $k_{dof}$ are 2, 4, 4, and 5 for MDA, MDB, MDC, and MDD, respectively. 
In Table~\ref{tab:bic}, we show the DTD model parameters which maximize the likelihood and their BIC values. 
The DTD profiles for these models are shown in Figure~\ref{fig:DTDcurve}. 

\begin{deluxetable*}{c|c|c|c|c|c|c|c}\label{tab:bic}
    \tablecaption{The BIC values for the 4 DTD models}
    \tablehead{
        \colhead{Model Name} & \colhead{$\tau$} & \colhead{$s$ or $s_1$} & \colhead{$s_2$} & \colhead{$r$} & \colhead{$t_c$} & \colhead{$k_{dof}$} & \colhead{BIC}
    }
    \startdata
    MDA & $10^{8.275}$ & -1.648 & None   & None         & None         & 2 & 495.49 \\
    MDB & $10^{8.320}$ & -2.363 & -1.167 & None         & $10^{9.263}$ & 4 & 503.68 \\ 
    MDC & $10^{8.385}$ & -2.566 & None   & $10^{1.069}$ & $10^{9.633}$ & 4 & 502.77 \\ 
    MDD & $10^{8.409}$ & -3.026 & -2.061 & $10^{1.268}$ & $10^{9.577}$ & 5 & 507.14 \\
    \enddata
    
\end{deluxetable*}

We notice that the one-component model MDA shows the smallest BIC value, while all the two-component models show larger BIC values. 
According to \citet{Robert1995BayesFactor}, the model with a smaller BIC value is preferred when the BIC value difference of the two models' is larger than 2. 
From this test, we conclude that given the current data set, we cannot establish the existence of a delayed component in the progenitors of SNe~Ia. 

\subsection{Cosmic Supernova Rate}\label{sec:csnr}

Given a cosmic SFH (CSFH), the cosmic SNe Ia rate (CSNR) in rest frame at different redshifts is a convolution of the DTD and the cosmic SFH: 

\begin{equation}
    CSNR(z)=DTD(z)\ast CSFH(z).
\end{equation}

We adopt the CSFH formula in rest frame from \citet{Madau2014CSFH}, which is
\begin{equation}
    CSFH(z)=0.015\frac{(1+z)^{2.7}}{1+[(1+z)/2.9]^{5.6}}\ M_{\odot}year^{-1}Mpc^{-3}.
\end{equation}

We use the observed CSNR data binned to redshift intervals from \citet{Strolger2020SNRates} to estimate the absolute SN rate $A$. The original SN rate data include those from \citet{Rodney2014Candels}, \citet{Rodney2010IfA}, \citet{Dahlen2008D08HST}, \citet{Graur2011sdf}, \citet{Graur2014Clash}, \citet{Perrett2012SNLS}, \citet{Okumura2014SubaruNewton}, \citet{Cappllaro2015vst}, \citet{Pain2002scp}, \citet{Neill2006SNLS}, \citet{Tonry2003highz}, \citet{Dilday2010SDSS}, \citet{Botticella2008stress}, \citet{Horesh2008SDSS}, \citet{Strolger2003PHD}, \citet{Madgwick2003SDSS}, \citet{Frohmaier2019PTF}, \citet{Mannucci2005C99}, \citet{Cappellaro1999C99}. 
All the errors (including statistic and systematic errors) in the observed CSNR are treated as Gaussian to calculate the normalization factor $A$ and its $1\sigma$ interval, using the the best-fit parameters in Table~\ref{tab:bic}. 
The results are shown in Table~\ref{tab:Apar}. 

\begin{deluxetable}{c|c}\label{tab:Apar}
    \tablecaption{The $A$ values for the 4 DTD models}
    \tablehead{
        \colhead{Model Name} & \colhead{$A (10^{-4} M_{\odot}^{-1})$}}
    \startdata
    MDA & $8.11_{-0.22}^{+0.23}$ \\
    MDB & $7.64_{-0.21}^{+0.21}$ \\ 
    MDC & $7.62_{-0.21}^{+0.20}$ \\ 
    MDD & $7.74_{-0.21}^{+0.21}$ \\
    \enddata
\end{deluxetable}

\begin{figure}[htb!]
    \plotone{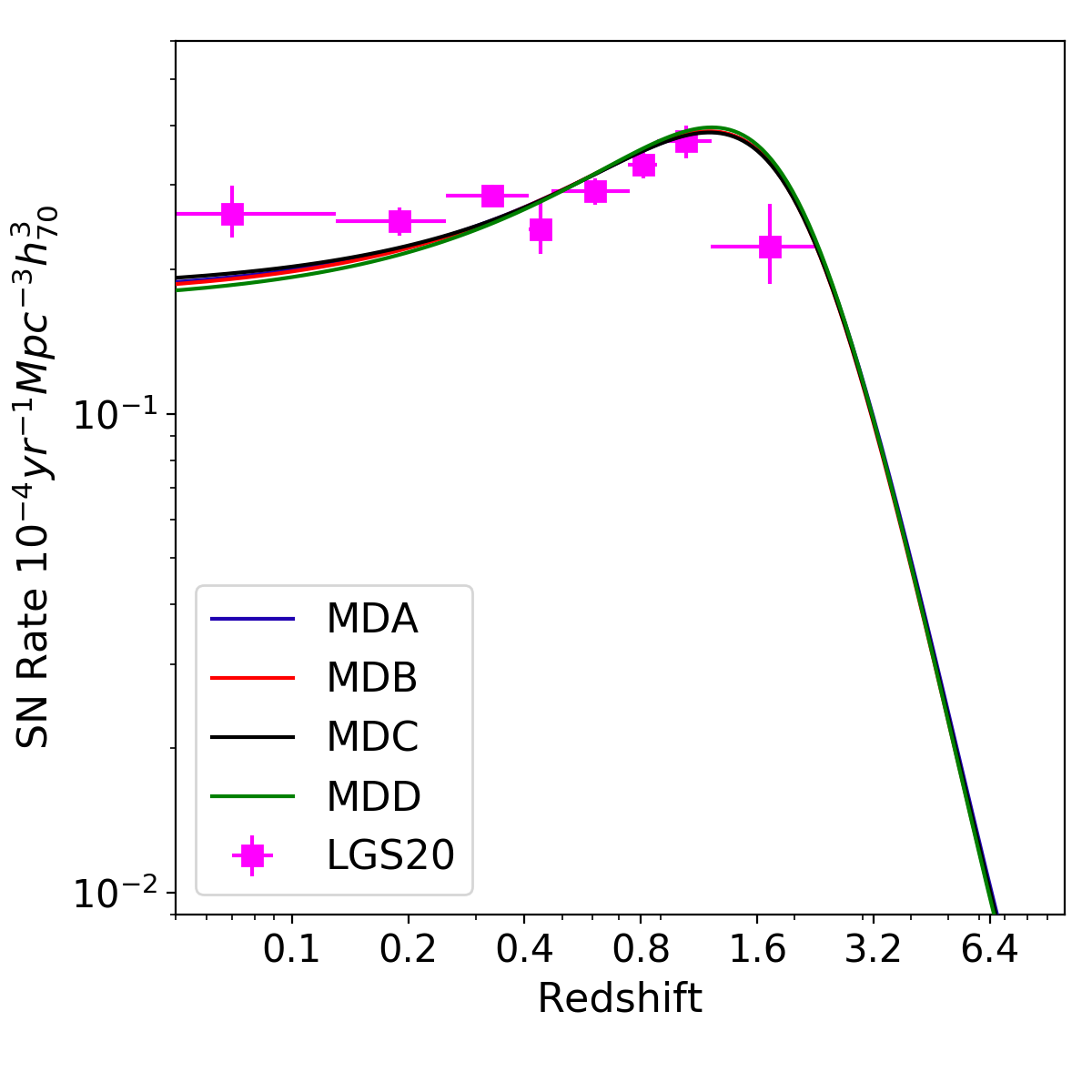}
    \caption{The CSNR for the DTD models MDA (ultramarine), MDB (red), MDC (black), MDD (green), and the binned observational CSNR from \citet{Strolger2020SNRates} (LGS20, magenta). The CSNRs in this figure are in observer frame. }\label{fig:CSNR}
\end{figure}

In Figure \ref{fig:CSNR}, we present the CSNR in observer frame derived from our four DTD models, and show the observed CSNR from \citet{Strolger2020SNRates}. 

\section{Conclusion}\label{sec:conclusion}

We selected 96 host galaxies of SNe~Ia, most of which are observed by the VLT+MUSE under the AMUSING program \footnote{https://amusing-muse.github.io/} 
to calculate the spatially resolved host galaxy SFHs and to constrain the DTD of the SN progenitors. 
A statistical method to spatially separate the SFH of a galaxy into multiple groups is developed and applied to constrain the model parameters of of the DTD models of SNe~Ia. 
We found the simple power law model MDA provides the best fit to the data, with the delay time  $\tau=120_{-83}^{+142}\ Myr$, and the SN rate decay slope $s=-1.41_{-0.33}^{+0.32}$. We have not found a significant delayed component of SN progenitors based on our analyses. 

Comparing to previous DTD estimates based on SN Ia rates at different redshifts (e.g. \citep{Matan2018z1d75}), our method does not rely on the details of SN search projects to deduce the redshift dependent SN rates. 

The slope parameter has been measured in previous studies. 
For example, \citet{Maoz2017Consistent} used a revised CSFH and derived SN Ia rates at different redshift bins up to $z\sim 2.25$ and constrained the slope to $s=-1.1_{-0.1}^{+0.1}$. 
\citet{Matan2018z1d75} used the HST to search for SNe~Ia in 12 massive galaxy clusters at $z\sim 1.13-1.75$, and measured the slope to be $s=-1.30_{-0.16}^{+0.23}$. 
\citet{Heringer2019DTD} used a color-luminosity method and an SFH reconstruction method on SDSS image survey data, and constrained the slope to be $s=-1.34_{-0.17}^{+0.19}$. 
In contrast, the delay time $\tau$ were measured with large uncertainties in previous studies. 
\citet{Maoz2010RateMeasure} measured the SN rates at redshift out to $z\sim 1.45$, and concluded that the delay time $\tau <2.2Gyr$. 
\citet{Maoz2010SNRMagCloud} analyzed SN~Ia remnants in the Magellanic Clouds, and found a ``prompt" SN Ia population, which explodes within 330 Myr of star formation. 
\citet{Maoz2012Sloan2} identified 90 SNe~Ia in SDSS II spectral survey program, and derived $\tau<0.42 Gyr$ from the SFHs of the SN host galaxies. 
Our result on $s$ is consistent with previous researches although with larger uncertainties, while our constraints on $\tau$ show a higher confidence level than previous studies. 
Note also, during the process of this paper, \citet{Castrillo2020Compete} showed  DTD constraints based on mean stellar age maps from {\tt\string Pipe3D} \citep{Sanchez2016pipe3d} using the same MUSE data set.
The results are $\tau=50_{-35}^{+100} Myr$ and $s=-1.1\pm 0.3$ (with 50\% confidence interval), which is consistent to our result to within the errors. 

When compared to the theoretical models of SNe~Ia, our DTD results prefer a DD scenario with CO WD+CO WD as progenitor system (e.g. \citep{Chen2012DD,LiuD2017WDHe}), which shows a delay time $\tau\sim 10^8 -10^{8.5} $ years and $s\sim -1$.
However, \citet{Ruiter2009Rates} and \citet{Mennekens2010TwoSim} predict a small ($\sim 5\%$, in \citet{Ruiter2009Rates}) fraction of ``prompt" SN Ia population is formed less than 100 Myr after star formation for the DD scenario. 
Due to the limitation of SFH code which can only  calculate the stellar population above 63 Myr, we did not attempt to introduce extra structure in our DTD models to accommodate this prompt population. 
Moreover, as discussed in many works  \citep[e.g.,][]{Mennekens2010TwoSim,LiuD2017WDHe}, the observed SN~Ia rates are most likely explained by a combination of two or more channels in the context of the BPS. 
Although our result indicates a high confidence of $\tau\sim 120 Myr$, we can not firmly establish or eliminate the dominant channels of SN~Ia progenitors. 

Several studies \citep[e.g.,][]{Perrett2012SNLS,Sullivan2006TwoComp}  suggested that SN~Ia DTD could have a two-component profile.
We did not find significant evidences for the two-component models in our research. 
This is likely due to the large SFH uncertainties with larger lookback time intrinsic to the SFH models we have employed. 

We have not derived strong constraints using the data from MaNGA. 
However, with the future data releases of SDSS-MaNGA survey program, a detailed analysis on the DTD may be worthwhile using the increased sample size. 

In summary, we have developed a new method to estimate the DTD of SNe~Ia which allows us to set constraints on the DTD model parameters. 
With data that we can expect from future observation programs with LSST \citep{Ivezi2019LSST}, DESI \citep{Aghamousa2016DESI}, and HETDEX \citep{Hill2008HETDEX}, we may expect this method to be applicable to other subtypes of SNe~Ia and core-collapse SNe to set strong constraints on the SN progenitors. 

\acknowledgements
Portions of this research were conducted with the advanced computing resources provided by Texas A\&M High Performance Research Computing. 
We thank transient name server website (\href{https://wis-tns.weizmann.ac.il}{https://wis-tns.weizmann.ac.il}) for providing the SNe data. 
This work is based on public data release from the MUSE commissioning observations at the VLT Yepun (UT4) telescope. 
We thank The All-weather MUse Supernova Integral field Nearby Galaxies (AMUSING) survey program \href{https://amusing-muse.github.io/}{https://amusing-muse.github.io/} especially, for their SN host galaxy survey proposals and their published data. 
XC would like to thank Prof. Casey Papovich, Prof. Jonelle Walsh, Dr. Yaswant Devarakonda, Dr. Jonanthan Cohn for supportive discussions. 
This paper has made use of the data from the SDSS projects. 
The SDSS-III web site is http://www.sdss3.org/. 
SDSS-III is managed by the Astrophysical Research Consortium for the Participating Institutions of the SDSS-III Collaboration. 
XC and LW are grateful to the support from an NSF grant AST 1817099.

\software{astropy \citep{astropy}, SExtractor \citep{sextractor}, ppxf \citep{ppxf2017}, sdss-marvin \citep{marvin}, emcee \citep{emcee}}

\clearpage

\appendix

\section{List of All Supernovae and Host Galaxies}\label{sec:snlist}

Table \ref{tab:museData} lists all the SNe and data information for the host galaxies observed by MUSE. Table \ref{tab:mangaData} lists all the SNe and data information for the host galaxies observed by MaNGA. 

\startlongtable
\begin{deluxetable*}{cccccccccc}
    \tablehead{\colhead{SN Name} & \colhead{SN RA} & \colhead{SN DEC} & \colhead{IFU RA} & \colhead{IFU DEC} & \colhead{SN Time} & \colhead{IFU Time} & \colhead{Redshift} & \colhead{ARCFILE} }
    \startdata
    SN2019fkq &  359.1010 & -29.0230 &  359.1016 & -29.0238 &  2019-05-14 &  2019-09-07 &    0.0450 &  ADP.2019-10-07T17:13:49.969 \\
    SN2018ezx &   62.0326 &  -8.8313 &   62.0337 &  -8.8332 &  2018-08-12 &  2016-11-08 &    0.0329 &  ADP.2017-01-18T15:19:35.833 \\
    SN2018djd &   33.6398 &  -0.7664 &   33.6410 &  -0.7658 &  2018-07-12 &  2017-10-28 &    0.0264 &  ADP.2017-11-20T17:51:27.822 \\
    SN2018zz &  210.9113 & -33.9780 &  210.9126 & -33.9786 &  2018-03-03 &  2015-08-08 &    0.0138 &  ADP.2016-07-12T07:52:45.162 \\
    SN2017hgz &  327.0808 & -34.9516 &  327.0838 & -34.9529 &  2017-10-10 &  2015-10-14 &    0.0162 &  ADP.2016-08-08T10:10:02.297 \\
    SN2017dps &  204.1639 & -33.9658 &  204.1668 & -33.9670 &  2017-05-01 &  2016-04-11 &    0.0125 &  ADP.2017-12-18T14:37:20.881 \\
    SN2017cze &  167.4450 & -13.3807 &  167.4451 & -13.3807 &  2017-04-11 &  2016-01-05 &    0.0149 &  ADP.2016-07-26T12:48:39.617 \\
    SN2016gfk &   18.5323 & -32.6519 &   18.5270 & -32.6572 &  2016-09-11 &  2016-05-20 &    0.0120 &  ADP.2016-10-05T16:09:44.597 \\
    SN2016aew &  212.8595 &   1.2867 &  212.8604 &   1.2860 &  2016-02-12 &  2014-06-24 &    0.0250 &  ADP.2016-08-02T10:23:14.733 \\
    SN2014dm &   62.0326 &  -8.8313 &   62.0297 &  -8.8270 &  2014-09-27 &  2016-11-08 &    0.0337 &  ADP.2017-01-18T15:19:35.833 \\
    SN2014at &  326.5628 & -46.5188 &  326.5618 & -46.5225 &  2014-04-20 &  2015-05-30 &    0.0325 &  ADP.2016-06-17T18:47:32.957 \\
    SN2014ao &  128.6391 &  -2.5461 &  128.6388 &  -2.5434 &  2014-04-17 &  2019-03-20 &    0.0139 &  ADP.2019-04-10T17:46:49.603 \\
    SN2013hk &   45.5462 &  15.9276 &   45.5452 &  15.9274 &  2013-12-04 &  2015-12-27 &    0.0170 &  ADP.2016-09-23T00:56:03.893 \\
    SN2013fz &   65.9446 & -51.5998 &   65.9435 & -51.5962 &  2013-11-02 &  2015-08-03 &    0.0206 &  ADP.2016-07-12T07:27:20.835 \\
    SN2013fy &  324.3678 & -47.0357 &  324.3630 & -47.0319 &  2013-10-25 &  2015-06-21 &    0.0314 &  ADP.2016-06-25T11:26:36.800 \\
    SN2013ef &   28.8417 &   6.6120 &   28.8363 &   6.6098 &  2013-07-04 &  2015-12-28 &    0.0172 &  ADP.2017-06-06T17:14:14.070 \\
    SN2013dl &   19.6732 &  -7.4494 &   19.6740 &  -7.4444 &  2013-06-17 &  2016-01-06 &    0.0024 &  ADP.2016-07-26T15:11:37.761 \\
    SN2013da &  206.4018 &  -7.3259 &  206.4009 &  -7.3257 &  2013-06-05 &  2017-04-01 &    0.0246 &  ADP.2017-04-11T12:41:12.920 \\
    SN2013az &   84.9729 & -40.5124 &   84.9672 & -40.5078 &  2013-03-24 &  2015-09-05 &    0.0373 &  ADP.2016-07-25T12:08:30.833 \\
    SN2013M &  209.9903 & -37.8637 &  209.9862 & -37.8637 &  2013-01-20 &  2017-04-18 &    0.0350 &  ADP.2017-12-12T14:16:55.449 \\
    SN2012he &   75.2111 & -38.6544 &   75.2086 & -38.6532 &  2012-11-22 &  2017-08-03 &    0.0576 &  ADP.2017-09-22T09:52:40.531 \\
    SN2012hd &   18.5323 & -32.6519 &   18.5311 & -32.6521 &  2012-11-20 &  2016-05-20 &    0.0120 &  ADP.2016-10-05T16:09:44.597 \\
    SN2012gm &  349.4017 &  14.0011 &  349.4043 &  14.0025 &  2012-11-19 &  2015-06-26 &    0.0148 &  ADP.2016-06-25T12:05:07.015 \\
    SN2012fw &  315.4961 & -48.2737 &  315.4958 & -48.2739 &  2012-08-19 &  2016-04-13 &    0.0186 &  ADP.2016-09-21T13:42:23.565 \\
    SN2012et &  355.6618 &  27.0922 &  355.6618 &  27.0921 &  2012-09-12 &  2016-06-09 &    0.0249 &  ADP.2016-09-29T20:36:01.636 \\
    SN2011jh &  191.8143 & -10.0621 &  191.8101 & -10.0631 &  2011-12-22 &  2019-02-20 &    0.0078 &  ADP.2019-03-07T06:28:16.831 \\
    SN2011iy &  197.2428 & -15.5177 &  197.2433 & -15.5178 &  2011-12-09 &  2016-05-12 &    0.0041 &  ADP.2016-09-29T05:21:54.104 \\
    SN2011iv &   54.7145 & -35.5881 &   54.7140 & -35.5922 &  2011-12-02 &  2017-11-22 &    0.0065 &  ADP.2017-12-13T01:47:07.213 \\
    SN2010jo &   14.3960 &  -1.3909 &   14.3982 &  -1.3926 &  2010-11-06 &  2017-07-20 &    0.0452 &  ADP.2017-09-11T14:28:03.988 \\
    SN2010ev &  156.3703 & -39.8282 &  156.3708 & -39.8309 &  2010-06-27 &  2016-04-13 &    0.0092 &  ADP.2016-09-21T13:42:23.507 \\
    SN2010dl &  323.7516 &  -0.5111 &  323.7540 &  -0.5133 &  2010-05-24 &  2017-08-04 &    0.0302 &  ADP.2017-09-22T10:41:54.300 \\
    SN2010aa &   27.1749 & -48.6480 &   27.1800 & -48.6502 &  2010-02-09 &  2018-05-26 &    0.0207 &  ADP.2018-06-02T02:30:18.655 \\
    SN2009jr &  306.6078 &   2.9102 &  306.6085 &   2.9092 &  2009-10-08 &  2017-07-30 &    0.0166 &  ADP.2017-09-20T13:08:51.796 \\
    SN2009iw &   88.8664 & -76.9201 &   88.8568 & -76.9211 &  2009-09-15 &  2015-09-24 &    0.0160 &  ADP.2016-07-28T11:30:23.655 \\
    SN2009fk &  341.1015 &  -0.1615 &  341.0996 &  -0.1617 &  2009-05-29 &  2017-08-04 &    0.0162 &  ADP.2017-09-22T10:41:54.308 \\
    SN2009ds &  177.2708 &  -9.7303 &  177.2671 &  -9.7291 &  2009-04-28 &  2016-06-30 &    0.0192 &  ADP.2017-10-16T10:25:08.202 \\
    SN2009aa &  170.9220 & -22.2711 &  170.9262 & -22.2707 &  2009-02-03 &  2015-04-07 &    0.0281 &  ADP.2016-06-09T16:16:30.539 \\
    SN2009Y &  220.5998 & -17.2527 &  220.5994 & -17.2468 &  2009-02-01 &  2016-04-03 &    0.0095 &  ADP.2016-09-07T10:11:23.531 \\
    SN2009I &   41.2915 &  -4.7106 &   41.2933 &  -4.7137 &  2009-01-13 &  2015-07-21 &    0.0262 &  ADP.2016-07-11T15:14:15.422 \\
    SN2008ia &  132.6464 & -61.2779 &  132.6465 & -61.2779 &  2008-12-07 &  2016-04-19 &    0.0217 &  ADP.2016-09-22T21:00:32.919 \\
    SN2008fu &   45.6195 & -24.4555 &   45.6188 & -24.4560 &  2008-09-25 &  2018-07-24 &    0.0524 &  ADP.2018-09-11T21:30:28.561 \\
    SN2008fl &  294.1897 & -37.5535 &  294.1868 & -37.5513 &  2008-09-07 &  2018-05-27 &    0.0199 &  ADP.2018-06-02T03:35:25.145 \\
    SN2008ec &  345.8151 &   8.8741 &  345.8190 &   8.8722 &  2008-07-14 &  2014-08-19 &    0.0159 &  ADP.2016-07-14T14:17:17.765 \\
    SN2008dh &    8.7973 &  23.2545 &    8.7972 &  23.2542 &  2008-06-08 &  2016-07-18 &    0.0368 &  ADP.2016-10-14T08:21:03.084 \\
    SN2008cf &  211.8831 & -26.5516 &  211.8857 & -26.5518 &  2008-05-04 &  2015-05-23 &    0.0471 &  ADP.2016-06-17T17:51:10.735 \\
    SN2008cc &  315.8740 & -67.1810 &  315.8734 & -67.1836 &  2008-04-24 &  2018-05-27 &    0.0106 &  ADP.2018-06-02T03:35:25.161 \\
    SN2008bq &  100.2658 & -38.0356 &  100.2605 & -38.0386 &  2008-04-02 &  2018-09-26 &    0.0346 &  ADP.2018-10-25T08:36:03.111 \\
    SN2008bd &  154.5978 & -13.1038 &  154.5972 & -13.1031 &  2008-03-13 &  2019-03-26 &    0.0306 &  ADP.2019-04-17T22:54:16.942 \\
    SN2008ar &  186.1585 &  10.8393 &  186.1580 &  10.8382 &  2008-02-27 &  2015-05-30 &    0.0262 &  ADP.2016-06-17T18:47:32.895 \\
    SN2007st &   27.1749 & -48.6480 &   27.1770 & -48.6494 &  2007-12-22 &  2018-05-26 &    0.0214 &  ADP.2018-06-02T02:30:18.655 \\
    SN2007so &   41.9318 &  13.2556 &   41.9297 &  13.2541 &  2007-12-13 &  2015-07-24 &    0.0298 &  ADP.2016-07-11T15:28:11.087 \\
    SN2007hx &   31.6127 &  -0.8992 &   31.6128 &  -0.8995 &  2007-09-03 &  2015-07-23 &    0.0798 &  ADP.2016-07-11T15:19:32.583 \\
    SN2007cq &  333.6697 &   5.0787 &  333.6685 &   5.0803 &  2007-06-21 &  2017-08-04 &    0.0263 &  ADP.2017-09-22T10:41:54.323 \\
    SN2007cg &  201.3917 & -24.6520 &  201.3899 & -24.6522 &  2007-05-11 &  2015-05-28 &    0.0331 &  ADP.2016-06-17T18:13:44.235 \\
    SN2007bc &  169.8142 &  20.8138 &  169.8107 &  20.8090 &  2007-04-04 &  2015-05-29 &    0.0208 &  ADP.2016-06-17T18:25:05.260 \\
    SN2007al &  149.8290 & -19.4729 &  149.8270 & -19.4738 &  2007-03-10 &  2015-05-31 &    0.0122 &  ADP.2017-03-28T14:09:36.373 \\
    SN2007ai &  243.2228 & -21.6266 &  243.2239 & -21.6302 &  2007-03-06 &  2018-05-26 &    0.0330 &  ADP.2018-06-02T02:30:18.663 \\
    SN2007S &  150.1291 &   4.4072 &  150.1302 &   4.4073 &  2007-01-29 &  2015-06-26 &    0.0139 &  ADP.2016-06-25T12:12:49.422 \\
    SN2006os &   43.7525 &  16.0126 &   43.7542 &  16.0097 &  2006-11-21 &  2015-07-21 &    0.0328 &  ADP.2016-07-11T15:14:15.414 \\
    SN2006ob &   27.9522 &   0.2636 &   27.9505 &   0.2634 &  2006-11-13 &  2015-07-13 &    0.0592 &  ADP.2016-07-11T14:04:30.018 \\
    SN2006lu &  138.8208 & -25.5999 &  138.8235 & -25.6001 &  2006-10-30 &  2015-04-13 &    0.0540 &  ADP.2016-06-14T09:15:58.860 \\
    SN2006hx &   18.4876 &   0.3719 &   18.4888 &   0.3717 &  2006-09-28 &  2015-06-21 &    0.0454 &  ADP.2016-06-25T11:26:36.852 \\
    SN2006hb &   75.5042 & -21.1342 &   75.5053 & -21.1320 &  2006-09-27 &  2018-08-29 &    0.0153 &  ADP.2018-10-17T14:54:22.224 \\
    SN2006et &   10.6911 & -23.5616 &   10.6909 & -23.5584 &  2006-09-03 &  2015-06-27 &    0.0223 &  ADP.2016-06-25T12:12:49.414 \\
    SN2006ej &    9.7512 &  -9.0149 &    9.7490 &  -9.0157 &  2006-08-23 &  2015-06-18 &    0.0203 &  ADP.2016-06-24T11:45:22.441 \\
    SN2006cm &  320.0731 &  -1.6842 &  320.0728 &  -1.6841 &  2006-05-24 &  2016-05-19 &    0.0163 &  ADP.2016-12-02T09:39:19.087 \\
    SN2006br &  202.5085 &  13.4164 &  202.5075 &  13.4158 &  2006-04-25 &  2015-06-04 &    0.0247 &  ADP.2016-06-24T10:20:34.507 \\
    SN2006D &  193.1445 &  -9.7772 &  193.1414 &  -9.7752 &  2006-01-11 &  2015-05-23 &    0.0086 &  ADP.2016-06-17T17:51:10.811 \\
    SN2005na &  105.4042 &  14.1366 &  105.4026 &  14.1332 &  2005-12-31 &  2015-04-11 &    0.0263 &  ADP.2016-06-21T00:31:05.284 \\
    SN2005lu &   39.0168 & -17.2638 &   39.0155 & -17.2639 &  2005-12-11 &  2015-06-18 &    0.0327 &  ADP.2016-06-24T11:45:22.429 \\
    SN2005ku &  344.9251 &  -0.0134 &  344.9275 &  -0.0137 &  2005-11-10 &  2015-05-30 &    0.0454 &  ADP.2016-06-17T18:47:32.872 \\
    SN2005iq &  359.6342 & -18.7111 &  359.6354 & -18.7092 &  2005-11-05 &  2018-07-25 &    0.0346 &  ADP.2018-09-13T01:03:07.237 \\
    SN2005hc &   29.2022 &  -0.2122 &   29.1998 &  -0.2137 &  2005-10-12 &  2015-08-02 &    0.0459 &  ADP.2016-07-12T07:19:14.082 \\
    SN2005bs &  302.5615 & -56.6390 &  302.5588 & -56.6454 &  2005-04-19 &  2016-05-13 &    0.0552 &  ADP.2016-09-29T08:33:33.612 \\
    SN2005bg &  184.3220 &  16.3717 &  184.3216 &  16.3716 &  2005-03-28 &  2015-05-29 &    0.0230 &  ADP.2016-06-17T18:47:32.927 \\
    SN2005be &  224.8876 &  16.6699 &  224.8863 &  16.6699 &  2005-04-05 &  2015-05-30 &    0.0336 &  ADP.2016-06-17T18:47:32.864 \\
    SN2005al &  207.5037 & -30.5772 &  207.5014 & -30.5762 &  2005-02-24 &  2018-05-10 &    0.0124 &  ADP.2018-05-18T04:03:44.301 \\
    SN2005ag &  224.1793 &   9.3286 &  224.1819 &   9.3285 &  2005-02-10 &  2015-05-29 &    0.0797 &  ADP.2016-06-17T18:25:05.248 \\
    SN2004gc &   80.4543 &   6.6794 &   80.4581 &   6.6760 &  2004-11-18 &  2019-02-24 &    0.0305 &  ADP.2019-03-08T05:11:20.934 \\
    SN2004ey &  327.2793 &   0.4473 &  327.2825 &   0.4442 &  2004-10-14 &  2015-06-05 &    0.0158 &  ADP.2016-06-24T10:28:35.821 \\
    SN2004ef &  340.5458 &  19.9971 &  340.5418 &  19.9946 &  2004-09-04 &  2015-06-05 &    0.0310 &  ADP.2016-06-24T10:28:35.837 \\
    SN2004do &  283.8988 & -53.7239 &  283.8905 & -53.7230 &  2004-08-04 &  2015-10-07 &    0.0086 &  ADP.2016-08-02T05:17:36.565 \\
    SN2004cs &  267.5579 &  14.2868 &  267.5599 &  14.2832 &  2004-06-23 &  2016-03-09 &    0.0141 &  ADP.2016-08-17T10:26:59.052 \\
    SN2003ic &   10.4605 &  -9.3035 &   10.4593 &  -9.3053 &  2003-09-16 &  2018-08-10 &    0.0554 &  ADP.2018-09-20T04:52:51.377 \\
    SN2003gh &  116.3256 & -71.4095 &  116.3247 & -71.4104 &  2003-06-29 &  2017-12-02 &    0.0179 &  ADP.2017-12-20T14:19:23.117 \\
    SN2002jg &  334.8675 &  29.3897 &  334.8700 &  29.3846 &  2002-11-23 &  2016-05-25 &    0.0162 &  ADP.2016-10-07T07:11:23.283 \\
    SN2002fk &   50.5283 & -15.3994 &   50.5238 & -15.4009 &  2002-09-17 &  2015-09-10 &    0.0071 &  ADP.2016-07-25T12:56:04.451 \\
    SN2001da &  358.3843 &   8.1183 &  358.3866 &   8.1174 &  2001-07-09 &  2016-05-22 &    0.0172 &  ADP.2016-10-05T16:52:03.490 \\
    SN2001E &  177.2708 &  -9.7303 &  177.2708 &  -9.7364 &  2001-01-05 &  2016-06-30 &    0.0192 &  ADP.2017-10-16T10:25:08.202 \\
    SN2001A &  184.8467 &   5.8251 &  184.8459 &   5.8279 &  2001-01-01 &  2016-04-17 &    0.0073 &  ADP.2016-09-22T13:48:58.454 \\
    SN2000fs &   47.1098 &   4.1109 &   47.1093 &   4.1111 &  2000-09-06 &  2018-11-03 &    0.0300 &  ADP.2018-11-12T14:39:55.596 \\
    SN2000do &  287.8565 & -50.6404 &  287.8591 & -50.6401 &  2000-09-30 &  2017-06-18 &    0.0109 &  ADP.2018-08-02T18:48:39.520 \\
    SN2000A &  351.9877 &   8.7785 &  351.9954 &   8.7839 &  2000-01-01 &  2017-09-15 &    0.0296 &  ADP.2017-10-06T15:09:34.514 \\
    SN1999ee &  334.0384 & -36.8439 &  334.0417 & -36.8444 &  1999-10-07 &  2014-10-27 &    0.0114 &  ADP.2016-06-23T09:51:35.962 \\
    SN1998V &  275.6593 &  15.6966 &  275.6558 &  15.7023 &  1998-03-10 &  2019-05-06 &    0.1753 &  ADP.2019-07-20T08:00:23.865 \\
    SN1997dt &  345.0154 &  15.9802 &  345.0122 &  15.9808 &  1997-11-22 &  2018-06-23 &    0.0073 &  ADP.2018-08-09T21:08:02.554 \\
    SN1994D &  188.5094 &   7.7007 &  188.5102 &   7.7013 &  1994-03-07 &  2016-05-26 &    0.0027 &  ADP.2016-10-06T16:52:10.915 \\
    SN1991S &  157.3645 &  22.0083 &  157.3658 &  22.0129 &  1991-04-10 &  2017-01-20 &    0.0544 &  ADP.2017-03-20T10:48:00.633 \\
    SN1989B &  170.0563 &  13.0061 &  170.0579 &  13.0053 &  1989-01-30 &  2018-05-14 &    0.0023 &  ADP.2018-05-29T18:17:29.607 \\
    SN1968I &  197.2036 &  -6.7776 &  197.2054 &  -6.7778 &  1968-04-23 &  2016-06-01 &    0.0056 &  ADP.2017-06-14T09:12:09.346 \\
    \enddata
    \tablecomments{SN Name: The names of SNe Ia. SN RA: Right ascension of SN coordinates. SN DEC: Declination of SN coordinates. IFU RA: Right ascension of IFU datacube center. IFU DEC: Declination of IFU datacube center. SN Time: SN discovery date. IFU Time: MUSE observation date. Redshift: SN host galaxy redshift. ARCFILE: The filename of IFU datacube product stored in ESO archive, to notice the timestamp is not the IFU observation time. }\label{tab:museData}
\end{deluxetable*}

\startlongtable
\begin{deluxetable*}{cccccccccc}
    \tablehead{\colhead{SN Name} & \colhead{SN RA} & \colhead{SN DEC} & \colhead{IFU RA} & \colhead{IFU DEC} & \colhead{SN Time} & \colhead{IFU Time} & \colhead{Redshift} & \colhead{IFU ID} }
    \startdata
    SN2018ccl &  247.0464 &  39.8201 &  247.0482 &  39.8219 &  2018-05-28 &  2015-06-23 &    0.0268 &  1-569169\\
    SN2018btb &  173.6162 &  46.3625 &  173.6187 &  46.3606 &  2018-05-14 &  2016-04-25 &    0.0338 &  1-279410 \\
    SN2018bbz &  261.9680 &  60.0961 &  261.9689 &  60.0973 &  2018-04-26 &  2015-09-04 &    0.0278 &   1-25680 \\
    SN2018ats &  153.2345 &  46.4181 &  153.2319 &  46.4177 &  2018-04-10 &  2015-03-25 &    0.0382 &  1-167380 \\
    SN2018aej &  236.0959 &  39.5581 &  236.0961 &  39.5590 &  2018-03-08 &  2015-06-11 &    0.0479 &  1-322806 \\
    SN2018ddh &  184.6835 &  44.7820 &  184.6847 &  44.7812 &  2018-07-01 &  2016-02-16 &    0.0383 &  1-258653 \\
    SN2017ckx &  117.0459 &  28.2303 &  117.0457 &  28.2303 &  2017-03-28 &  2015-11-12 &    0.0272 &  1-556501 \\
    SN2012hj &  166.8300 &  46.3795 &  166.8320 &  46.3833 &  2012-12-04 &  2015-02-14 &    0.0246 &  1-277539 \\
    SN2012bm &  196.4402 &  46.4647 &  196.4440 &  46.4619 &  2012-03-27 &  2015-05-08 &    0.0248 &  1-284329 \\
    SN2007sw &  183.4037 &  46.4934 &  183.4036 &  46.4939 &  2007-12-29 &  2015-03-15 &    0.0257 &  1-575847 \\
    SN2007R &  116.6564 &  44.7895 &  116.6571 &  44.7905 &  2007-01-26 &  2014-10-29 &    0.0308 &  1-339041 \\
    SN2006iq &  324.8906 &  10.4849 &  324.8916 &  10.4835 &  2006-09-23 &  2014-08-31 &    0.0789 &  1-114465 \\
    SN2006cq &  201.1046 &  30.9563 &  201.1059 &  30.9593 &  2006-05-29 &  2017-06-15 &    0.0485 &  1-575232 \\
    SN2003an &  201.9731 &  28.5081 &  201.9719 &  28.5082 &  2003-02-09 &  2017-03-02 &    0.0370 &  1-395622 \\
    SN2002aw &  249.3711 &  40.8806 &  249.3720 &  40.8799 &  2002-02-15 &  2016-03-16 &    0.0264 &  1-135668 \\
    SN2002G &  196.9803 &  34.0851 &  196.9784 &  34.0871 &  2002-01-18 &  2017-05-18 &    0.0336 &  1-415476 \\
    SN2004H &  173.4990 &  49.0629 &  173.4968 &  49.0620 &  2004-01-17 &  2017-04-16 &    0.0316 &  1-576106 \\
    PTF11bui &  198.2350 &  47.4535 &  198.2363 &  47.4566 &  2011-04-26 &  2015-04-15 &    0.0281 &  1-285004 \\
    PTF11mty &  323.5217 &  10.4235 &  323.5212 &  10.4219 &  2011-09-23 &  2014-08-31 &    0.0774 &  1-114129 \\
    PTF12izc &  355.8932 &   0.5687 &  355.8949 &   0.5678 &  2012-09-21 &  2015-09-16 &    0.0826 &   1-29726 \\
    PTF13f &  247.3591 &  38.4198 &  247.3615 &  38.4194 &  2013-02-01 &  2015-06-23 &    0.0305 &  1-211264 \\
    Gaia15abd &  205.2830 &  23.2830 &  205.2827 &  23.2821 &  2015-02-07 &  2017-03-01 &    0.0264 &  1-568584 \\
    SN2017frb &  317.9036 &  11.4974 &  317.9032 &  11.4969 &  2017-07-25 &  2014-07-04 &    0.0294 &  1-113540 \\
    SN2019pig &  225.3894 &  49.1095 &  225.3890 &  49.1124 &  2019-09-03 &  2016-04-28 &    0.0260 &  1-246549 \\
    SN2019omi &   57.2491 &   0.9269 &   57.2484 &   0.9260 &  2019-08-24 &  2015-11-06 &    0.0358 &  1-229060 \\
    SN2017fel &  322.3057 &  -0.2947 &  322.3060 &  -0.2948 &  2017-07-05 &  2015-09-13 &    0.0305 &  1-289846 \\
    SN2010dl &  323.7540 &  -0.5133 &  323.7516 &  -0.5114 &  2010-05-24 &  2016-06-15 &    0.0302 &  1-180080 \\
    SN2007O &  224.0216 &  45.4047 &  224.0182 &  45.4053 &  2007-01-21 &  2017-03-04 &    0.0362 &  1-576436 \\
    SN2006np &   46.6645 &   0.0640 &   46.6649 &   0.0620 &  2006-11-10 &  2016-11-01 &    0.1074 &   1-37863 \\
    SN2002ci &  243.9081 &  31.3215 &  243.9074 &  31.3213 &  2002-04-19 &  2016-05-14 &    0.0222 &  1-272321 \\
    \enddata\label{tab:mangaData}
\end{deluxetable*}
    
\clearpage


\bibliographystyle{aasjournal}

\end{document}